\documentclass[%
 reprint,
 amsmath,amssymb,
 aps,
]{revtex4-1}

\usepackage{graphicx}
\usepackage{dcolumn}
\usepackage{bm}
\usepackage{hyperref}
\usepackage{comment}
\usepackage{subfigure}
\usepackage{mathtools}
\usepackage{bigints}
\usepackage{longtable}
\usepackage{graphicx} 
\usepackage{float}
\usepackage{empheq}
\usepackage{nccmath}
\usepackage{bm}

\begin{document}


\title{Dynamics and stability of a compound particle - a theoretical study}

\author{Sravana Chaithanya}
 \altaffiliation{Department of Chemical Engineering, Indian Institute of Technology Madras, Chennai 600036, India. E-mail: sumesh@iitm.ac.in}
\author{Sumesh P Thampi}





\begin{abstract}
Particles confined in droplets are called compound particles. They are encountered in various biological and soft matter systems. Hydrodynamics can play a decisive role in determining the configuration and stability of these multiphase structures during their preparation and use. Therefore, we investigate the dynamics and stability of a concentric compound particle under external forces and imposed flows. Governing equations are solved analytically in the inertia-less limit using the standard technique of superposition of vector harmonics and the solutions obtained are reported in terms of steady state flow fields, viscous drag on the particle and the time evolution of the confining drop shape. The limiting form of compound particle as a thin film coated rigid particle is analyzed in each case. We find that concentric configuration of a rotating compound particle is a steady state solution, and we calculate the extra force required to stabilize the concentric configuration of a translating compound particle. A comprehensive comparison of drop deformations in various linear ambient flows is also provided. Based on the findings, we propose pulsatile flow as a reliable method to transport compound particles without breakup of the confining drop. Thus, our analysis provides useful guidelines in preparation and transportation of stable compound particles in the context of nucleated cells, aerosols, droplet-based encapsulation of motile organisms and polymer microcapsules.
\end{abstract}

\pacs{Valid PACS appear here}
\maketitle


A solid particle confined in a droplet is termed as a compound particle. Such multiphase structures are ubiquitous in biological and soft matter systems. For example, cells encapsulated in polymeric fluid drops are  prepared and used commonly in tissue engineering \cite{cell-polymer-1,cell-polymer-2}.  In the field of genetic and developmental research of organisms it is desirable to focus on an individual and control its microenvironment - two aims easily achievable by confining it in droplets \cite{Wen,Zhang,Plessy,Lauga}. Compound particles have recently been encountered in self cleaning super hydrophobic surfaces \cite{Wisdom} where the liquid drops enclose the contaminants. Compound particles have also been used to control the equilibrium morphology of double emulsions\cite{control_morphology}. Colloidal particles with a core-shell structure is yet another example where compound particles are generated as an intermediate\cite{microcapsule}. In all the applications, the morphology of the compound particle namely the  spatial distribution of the solid inclusion as well as the droplet shape and the stability of the confining drop against breakup are dependent on the production and processing conditions. Therefore, in this work, we study theoretically the effect of external forces and surrounding fluid flows on the dynamics of a concentric compound particle and investigate the stability of this configuration.

The complexity of compound particle dynamics arises from the hydrodynamic interaction between the solid inclusion and the confining fluid interface \cite{JhonsonARFM}. One consequence of this interaction is that the inclusion and the confining drop surface may have different  velocities. Many of the previous studies consider compound droplets where the inclusion is a fluid. Dynamics of translating compound droplets for concentrically \cite{Rushton} and eccentrically \cite{Sadhal_eccentric,Qu} located fluid inclusions as well as inclusions partly covered with thin liquid films \cite{Sadhal} have been examined in the literature. The confining fluid surface may also be deformed in presence of an externally imposed flow. However, this deformation may be different from that of a simple drop, again due to the hydrodynamic interaction between the interface and the inclusion in a compound drop. The deformation characteristics of the confining interface when subjected to a linear flow has been investigated in the literature both analytically and numerically \cite{Brenner,lealstone}. 
Following these works, studies have been extended now in various directions for example, to incorporate the role of fluid inertia \cite{bazhlekov1995unsteady} and viscoelasticity \cite{zhou2006formation}, the effect of surfactant laden interfaces \cite{surfactant}, confinement \cite{confinement, halim},  and charged droplets which respond to electric fields \cite{charged}.

However, most of the classic studies \cite{JhonsonARFM, Rushton, Sadhal, Brenner, lealstone} mentioned above were inspired by and therefore concentrated on compound droplets. While this generalized approach encompasses compound particles as a subset, the properties of the enclosed fluid drop and its deformation characteristics increase the number of variables to be studied in the problem and making the analysis cumbersome. Therefore an extensive analysis on the effect of viscosity of fluids and size of inclusions on the morphology and configurational stability specific to a compound particle is desirable. This is an important aspect in many recent developments like  encapsulation of cells, organisms, contaminants or colloidal particles. Therefore we revisit and expand the study of the dynamics of a compound particle driven by either external forces or subjected to external flows. Since compound particles typically encounter variety of fluid flows as it moves through the junctions and length of a microfluidic channel we compare the response of the compound particle to canonical flows such as simple shear, uniaxial and biaxial flows. Further, we expand the scope of the work by investigating the unsteady evolution of the confining drop shape and propose a method to increase the structural stability of compound particles when subjected to external flows. 

This paper is organized as follows. We first describe the problem statement and provide the governing equations and boundary conditions. The solutions for the case of a rotating and translating compound particle in response to an externally imposed torque and force respectively are discussed then. The response of a compound particle to a linear flow, deformation dynamics of the confining interface, and stability characteristics of these compound particles in response to various externally imposed flows are analyzed. Finally we conclude by proposing pulsatile flow as a method to increase the stability of a compound particle during any process operations.
 
\section{Problem statement}
\begin{figure}[b]
\centering
  \includegraphics[height=5.0cm]{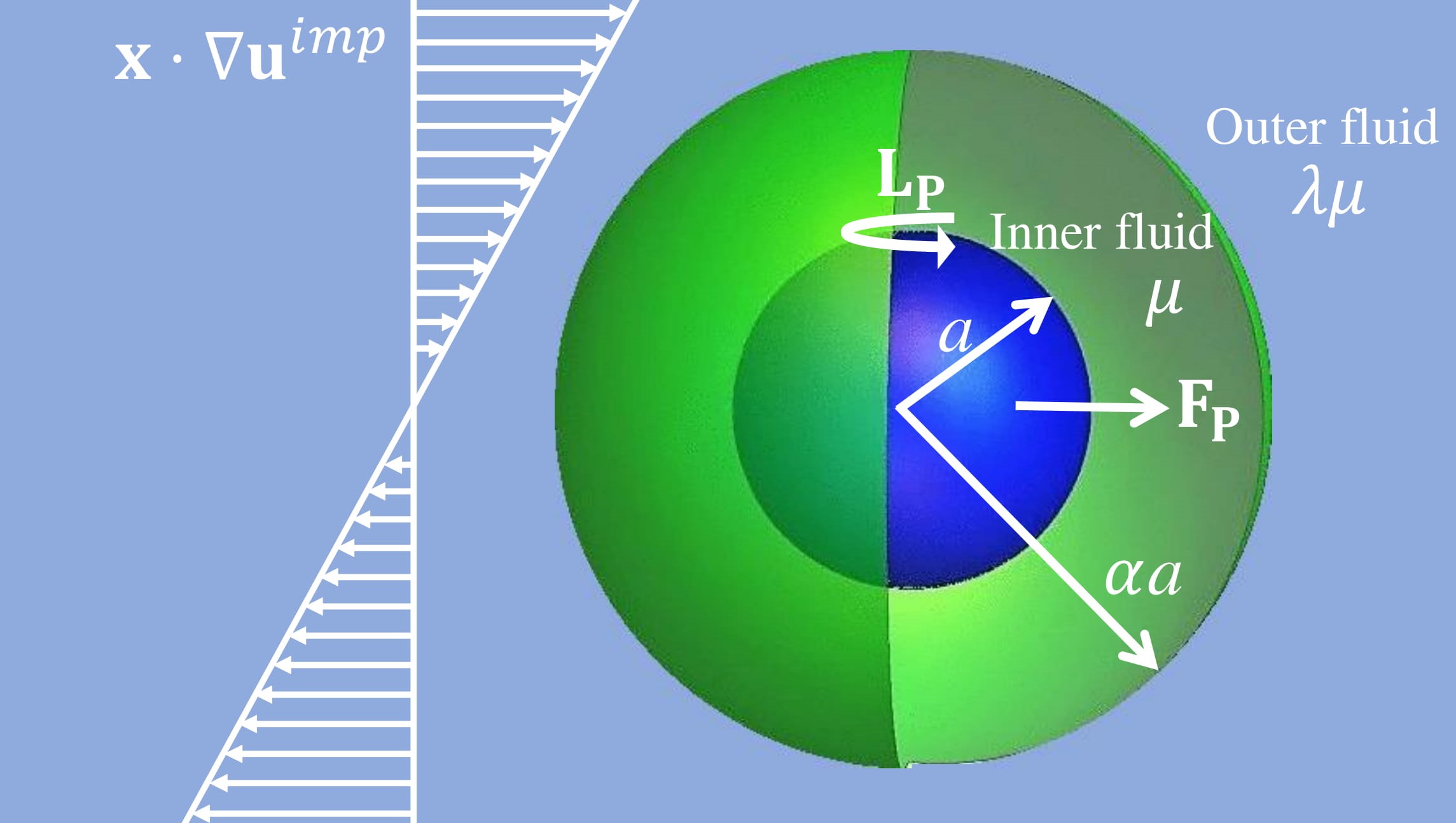}
   \caption{Schematic of the compound particle system. A solid particle of radius $a$ enclosed in a fluid drop of radius $\alpha a$ and viscosity $\mu$ (inner fluid), dispersed in another fluid of viscosity $\lambda \mu$ (outer fluid). It is subjected to an external force $\mathbf{F}_{P}$, torque $\mathbf{L}_{P}$ and an imposed flow field $\mathbf{x} \cdot \nabla \mathbf{u}^{imp}$.}
      \label{fig:schematic}
\end{figure}
Consider a compound particle having a spherical solid inclusion of radius $a$ placed concentrically in a drop of radius $\alpha a$  as shown in Fig~\ref{fig:schematic}. The inner and outer fluids are assumed to be Newtonian fluids of viscosity $\mu$ and $\lambda \mu$ respectively. An external force, $\mathbf{F}_{P}$, torque $\mathbf{L}_{P}$ are acting on the compound particle which is also subjected to an imposed linear flow $\mathbf{x} \cdot \nabla \mathbf{u}^{imp}$. Depending upon the form of $\nabla\mathbf{u}^{imp}$, different flow fields may be obtained as discussed later in section~\ref{sec:linearflow}.

We assume that the Reynolds number is very small, indicating that viscous forces are much larger than the inertial forces\cite{russel}. This is justified because, in typical applications, the size of compound particles may vary between 1-100$\mu$m.
Then the governing equations for the fluid flow are given by steady Stokes' equations \cite{leal2007advanced}
\begin{equation}
    \nabla \cdot \mathbf{u}^{(k)}= 0, \quad   \nabla \cdot \mathbf{T}^{(k)}= 0,
    \label{AT-gov-eqs}
\end{equation}
where $\mathbf{u}^{(k)}$ is the velocity field in each phase ($k=1, \ 2$), $\mathbf{T}^{(k)} = -p^{(k)}\mathbf{I}+\mu^{(k)}[\nabla \mathbf{u}^{(k)}+(\nabla \mathbf{u}^{(k)})^T]$ is the stress field and $p^{(k)}$ is the dynamic pressure.

The following boundary conditions are used. On the particle surface, at $\mathbf{x} = \mathbf{x}_s$ we have $\mathbf{u}^{(1)} = \mathbf{V}_{P} + \boldsymbol{\Omega}_P \wedge \mathbf{x}_s - \mathbf{x}_s \cdot \nabla \mathbf{u}^{imp} $ where, $\mathbf{V}_P$ and $\boldsymbol{\Omega}_P$ are respectively the translational and rotational velocity of the particle. Similarly the drop velocities are prescribed by $\mathbf{V}_D$ and $\boldsymbol{\Omega}_D$. In addition, continuity of velocity, $\mathbf{u}^{(1)} = \mathbf{u}^{(2)}$ and stress balance, 
 \begin{equation}
 \mathbf{T}^{(2)}\cdot \mathbf{n} - \mathbf{T}^{(1)}\cdot \mathbf{n} = \gamma (\nabla \cdot \mathbf{n})\mathbf{n}.
 \label{AT-stress-continuity}
 \end{equation}
are maintained at the interface. Here $\gamma$ is the interfacial tension and $\mathbf{n}$ represents unit normal to the interface. The relevance of capillary stress can be gauged by defining capillary number $Ca = {\lambda \mu {V}_{D}}/{\gamma}$. 
Defining a scalar function $S(\mathbf{x,t}) = 0$ on the interface, the kinematic condition on the interface is written down as \cite{leal2007advanced}
\begin{equation}
    \frac{1}{|\nabla S|} \frac{\partial S}{\partial t}+Ca (\text{$\mathbf{u}^{(I)}$}\cdot\mathbf{n}) = 0,
    \label{LF-kinematic-condition}
\end{equation}
where $\mathbf{u}^{(I)}$ is the velocity evaluated at the interface.

Though the governing equations are linear, the problem is non-linear because the shape of the droplet is an unknown variable coupled to the flow field. To make the problem analytically solvable, we assume that the droplet deformations are small, namely $Ca<<1$. Then all variables are expanded in $Ca$, an approach proved useful in studying emulsions \cite{Acrivos}. We solve the flow problem at the leading order ($O(1)$) and simultaneously determining the interface deformation at $O(Ca)$. Since the problem remains linear at the leading order, we divide the problem into three sub-problems and discuss the solutions of each of them below. Each problem is solved using the standard technique of superposition of vector harmonic functions as a solution to Laplace's equation \cite{stone_quadratic}.

 \section{Results and discussion}
\subsection{Rotation of a compound particle} 

Consider the first case, a rotating compound particle. Due to the imposed torque, if the angular velocity of the solid inclusion is $\boldsymbol{\Omega}_P$ then the solution to Eq.~\ref{AT-gov-eqs} describing the flow field may be obtained as,
\begin{align}
\mathbf{u}^{(1)} &= \frac{(1-\lambda)}{1+\lambda(\alpha^3-1)}\boldsymbol{\Omega}_{P}\wedge\mathbf{x}+\frac{\lambda a^{3} \alpha^{3} }{ 1 + \lambda ( \alpha^3 - 1 ) } \boldsymbol{\Omega}_{P}\wedge\frac{\mathbf{x}}{r^{3}}, \\ \quad p^{(1)} = 0,\nonumber \hspace{-1cm}\\
    \mathbf{u}^{(2)} &= \frac{a^{3}\alpha^{3}}{1+\lambda(\alpha^3-1)}\boldsymbol{\Omega}_{P}\wedge\frac{\mathbf{x}}{r^{3}}, \quad p^{(2)} = 0.\nonumber
    \label{RD-u2-complete}
\end{align}
where $r = |\mathbf{x}|$. Velocity fields of both inner and outer fluids have no radial component and are purely azimuthal. 

\begin{figure}[t]
\centering
\subfigure[]{\includegraphics[height=5 cm]{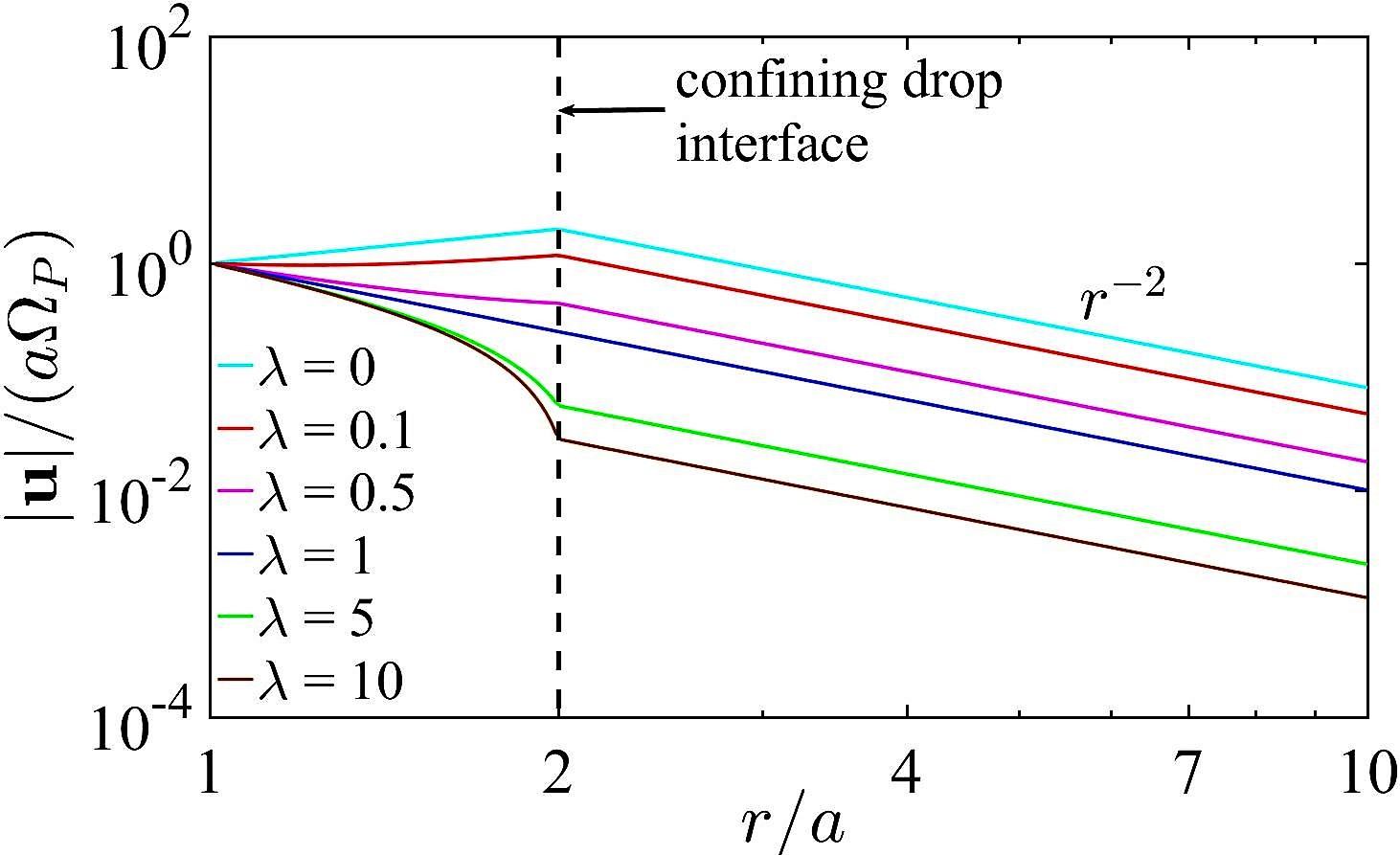}
\label{fig:RD_vel_decay}}
\subfigure[]{\includegraphics[height=6.5  cm]{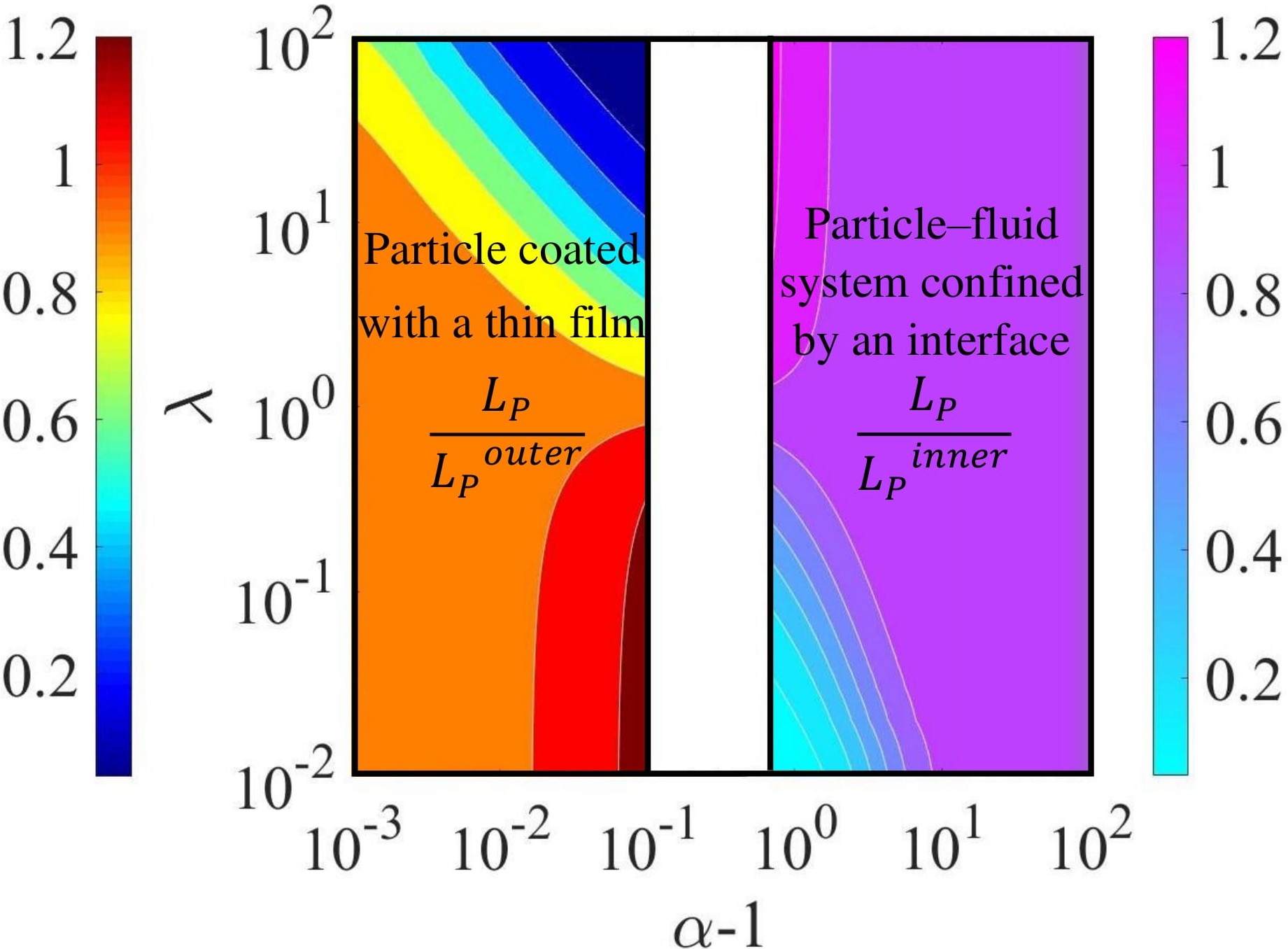}    \label{fig:RD_torque}}
  \caption{Velocity field and torque calculated due to rotation of a compound particle (a) Change in the azimuthal velocity as a function of radial distance from the centre of the solid particle for $\alpha = 2$ and various $\lambda$. (b) Normalized torque on a particle rotating inside a drop as a function of the thickness of the confining drop ($\alpha-1$) and viscosity ratio ($\lambda$). The contour plot on the left hand side shows the modified torque on a particle due to a thin film coating. The plot on the right hand side shows the modified torque on a particle when particle-fluid system is confined by an interface.}
\end{figure}

The spatial variation of the azimuthal velocity in both the fluids is plotted in Fig.~\ref{fig:RD_vel_decay}. When viscosity of the inner fluid is same as that of the outer fluid, the system behaves as if it is a single fluid and velocity decays as $r^{-2}$. As the viscosity of the inner fluid decreases ($\lambda \to \infty$), this decay is much faster inside the drop. On the other hand, if the inner fluid has a larger viscosity ($\lambda \to 0$), velocity of the inner fluid increases with distance (as in solid body rotation) inside the drop before it decays further outside the drop.

Since there is no radial component of fluid velocity and the dynamic pressure is zero, the drop does not deform and maintains the spherical shape. Angular velocity of the drop is calculated as
\begin{equation}
   \boldsymbol{\Omega}_{D} = \frac{1}{1+\lambda(\alpha^3-1)}\boldsymbol{\Omega}_P.
   \label{RD-Omega-drop}
\end{equation}
Thus the interface always has an angular velocity smaller than that of the particle for any $\lambda \neq 0$. Even though the angular velocity of the drop and the particle are different, the concentric configuration of the compound particle is maintained since rotation does not change the configuration.

The viscous torque experienced by the rotating particle can be calculated from the stress field ($\mathbf{T}$) on the surface of the particle as
\begin{equation}
    \mathbf{L}_{P} = -8\pi\boldsymbol{\Omega}_{P}\mu a^{3} \frac{\lambda  \alpha^{3}}{1+\lambda(\alpha^3 - 1)} . 
    \label{RD-Torque}
\end{equation}
As illustrated in Fig.~\ref{fig:RD_torque}, this expression gives solutions of two interesting cases.
\begin{enumerate}
\item \textit{Modified torque on a particle due to a thin film coating:} Consider a rotating particle. If the surrounding fluid is only the outer fluid then the viscous torque experienced by the particle is given by $\mathbf{L}_{P}^{outer} = -8\pi \boldsymbol{\Omega}_{P} \lambda \mu a^{3} $. This drag gets modified if the particle surface is covered by a coating of another fluid, say the inner fluid. If the thickness of the coating fluid  $\alpha - 1 << 1$ then the modified torque due to the presence of coating is,
\begin{equation}
\mathbf{L}_{P} = \mathbf{L}_{P}^{outer} \Big(1 + 3 (1- \lambda) (\alpha - 1) \Big)+O(\alpha-1)^{2}.
\label{RD-thin-film-torque} 
\end{equation}
If the viscosity of the coating fluid is very large ($\lambda \to 0$), then the drag on the particle increases and this modification to the  torque depends only on the film thickness $\alpha - 1$ and not on the viscosity of the coating fluid itself. On the other hand, if the viscosity of the coating fluid is small then the drag on the particle decreases and this modification to the torque is a linear function of both coating thickness $(\alpha-1)$ and viscosity ratio $\lambda$. Both the changes and the contrast in the dependence of viscosity ratio on the drag on a film coated particle are apparent in the left hand side of Fig.~\ref{fig:RD_torque}.
 
\item \textit{Modified torque on a particle when particle-fluid system is confined by an interface:}  The viscous torque experienced by a particle rotating in the inner fluid alone is given by $\mathbf{L}_{P}^{inner} = -8\pi \boldsymbol{\Omega}_{P} \mu a^{3} $. This drag gets modified if the system containing the particle and the inner fluid is confined by an outer fluid. Right hand side of Fig.~\ref{fig:RD_torque} shows this modification to the torque. If the viscosity of the outer fluid is very large ($\lambda \to \infty$) the torque on the particle increases since the motion of the inner fluid is hindered by the outer fluid. On the other hand if the viscosity of the outer fluid is very small, the inner fluid can slip past the outer fluid thus remarkably reducing the viscous torque experienced by the particle. Such modifications are significant when the size of the particle is comparable to the size of the drop. As the size of the drop increases, effect of confinement decreases and the torque approaches $\mathbf{L}_{P}^{inner}$. 
\end{enumerate}

 \subsection{Translation of a compound particle}
 In this section, we consider the second problem, the case of a translating compound particle under the action of an external force.
 
If the solid inclusion is translating with a velocity $\mathbf{V}_P$ we obtain the velocity and pressure fields as,
\begin{equation}
\begin{split}
 u_{i}^{(1)} &= c_{i}+m_{j}M_{ij}+g_{j}^{(1)}G_{ij}+d_{j}^{(1)}D_{ij}, 
 \end{split}
\end{equation}
\begin{equation}
\begin{split}
\quad p^{(1)} = 10 m_j x_j +2\frac{g_{j}^{(1)} x_j}{r^3},\nonumber 
\end{split}
\end{equation}
\begin{equation}
\begin{split}
 u_{i}^{(2)} &= g_{j}^{(2)}G_{ij}+d_{j}^{(2)}D_{ij}, \quad p^{(2)} = 2\frac{g_{j}^{(2)} x_j}{r^3}, \nonumber
\end{split}
\end{equation}
where $M_{ij} = 2r^2\delta_{ij}-x_ix_j$ represents the stokeson, $G_{ij} =  \frac{\delta_{ij}}{r}+\frac{x_ix_j}{r^3}$ the stokeslet, $D_{ij} = -\frac{\delta_{ij}}{r^3}+3 \ \frac{x_ix_j}{r^5}$ the potential force dipole \cite{Pozrikidis1992}. The expressions for the coefficients $c_{i}$, $m_{j}$, $g_{j}^{(k)}$, and $d_{j}^{(k)}$ are given in the appendix A.

 \begin{figure}[]
 \centering
 \subfigure[]{
  \includegraphics[height=3.8cm]{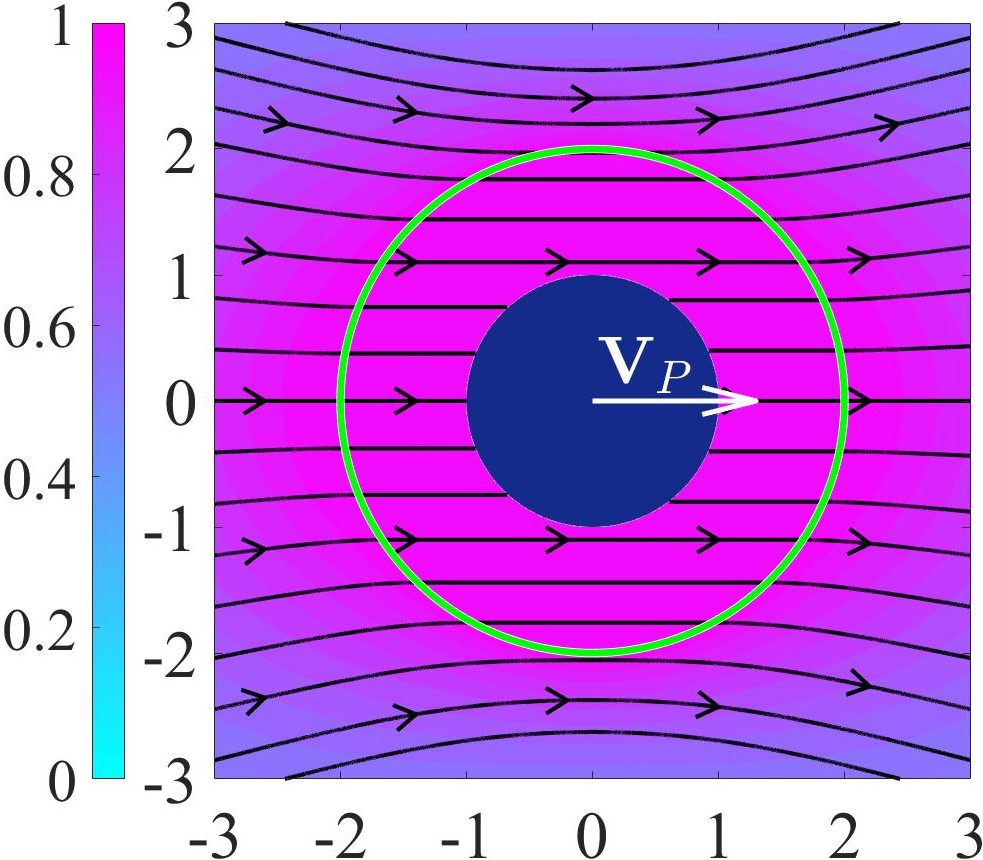}
  \label{fig:TD_vel_contour_0}}
 \subfigure[]{
  \includegraphics[height=3.8cm]{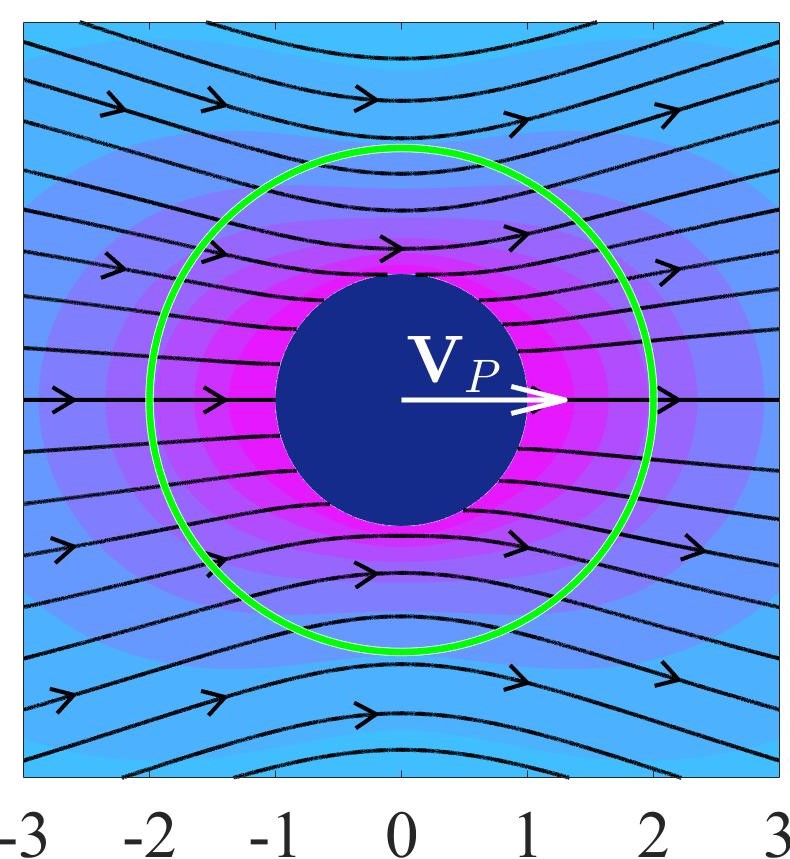}
   \label{fig:TD_vel_contour_1}} 
  \subfigure[]{
   \includegraphics[height=3.8cm]{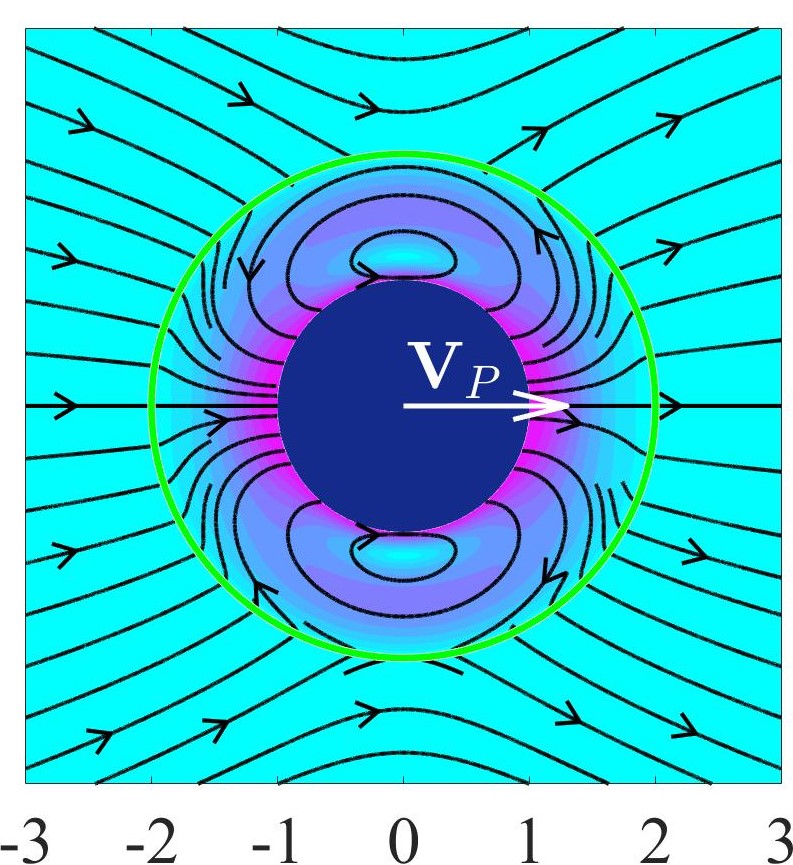}
   \label{fig:TD_vel_contour_100}}
      \subfigure[]{
  \includegraphics[height=3.8cm]{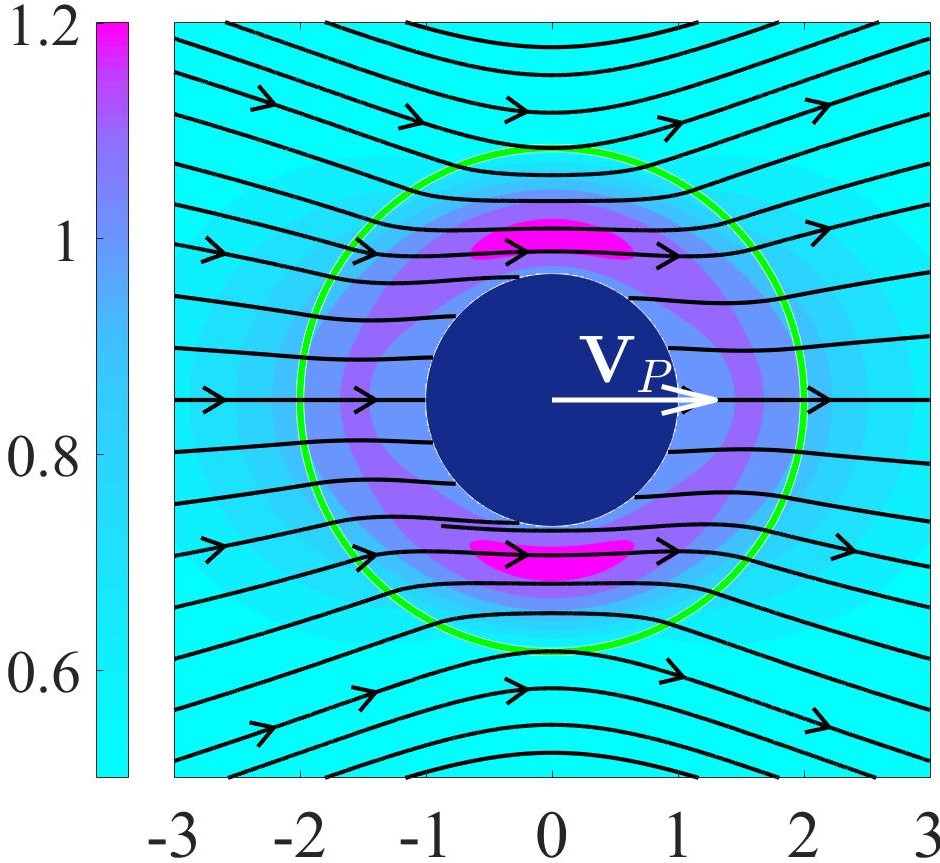}
       \label{fig:TD_stable_vel_contour_100}}
    \caption {Flow field generated by the particle translating inside a droplet for $\alpha = 2$, (a) $\lambda = 0$, (b) $\lambda = 1$, (c) $\lambda = 100$. The domain is colored according to the magnitude of velocity $|\mathbf{u}|/V_P$, on which streamlines (continuous lines) are superimposed. (d) Velocity field around a stable concentric configuration of a translating compound particle ($\alpha = 2,~\lambda = 100$).}
\label{fig:translation}
 \end{figure}

The typical velocity fields generated by a translating compound particle is illustrated in Fig.~\ref{fig:translation}. When the viscosity of the inner and outer fluids are same, the velocity field is similar to that of a translating rigid particle in a single fluid (Fig.~\ref{fig:TD_vel_contour_1}). 
If the viscosity of the outer fluid is much smaller than that of the inner fluid, mobility of the interface increases (Fig.~\ref{fig:TD_vel_contour_0}) and consequently the
velocity of the inner fluid approaches $\mathbf{V}_{P}$. On the other hand if the outer fluid has much larger viscosity than the inner fluid, then the interface motion is retarded. Consequently, the inner fluid recirculates (Fig.~\ref{fig:TD_vel_contour_100}). Generation of this recirculating flow depends upon both $\lambda$ and $\alpha$. For very small $\alpha$, large viscosity contrast is needed for generating recirculations but for large $\alpha$ recirculating flow fields are observed for $\lambda > 5$. Such recirculating flow fields may be useful to enhance mixing and thus improve transport properties in a compound particle system.

The generated velocity and pressure fields satisfy the boundary conditions on the confining spherical interface exactly and thus the confining drop remains undeformed in this case as well. 

Similar to the viscous torque discussed in the previous section, calculation of the drag force on the particle also gives the solution to two interesting cases.

\begin{enumerate}
 \item \textit{Modified Stokes drag on a particle due to a thin film coating:}
\begin{equation}
    \mathbf{F}_{P} = \mathbf{F}_{P}^{outer} \Big(1+ (1-\lambda)(\alpha-1)\Big)+O(\alpha-1)^2,
    \label{eqn:TD_unstable_thin_film_drag_force}
\end{equation}
where $F_{P}^{outer} =-6 \pi \lambda \mu a \mathbf{V}_{P}$ and

\item\textit{Modified Stokes drag on a particle when the particle-fluid system is confined by an interface:}
\begin{equation}
    \mathbf{F}_{P} = \mathbf{F}_{P}^{inner} \Bigg(\lambda \alpha  \frac{4(1-\lambda) + 2 \alpha^{5}(3+2\lambda)}{4+6\alpha^5+\lambda Z_1+\lambda^2 Z_2}\Bigg),
    \label{eqn:TD_unstable_drag_force}
\end{equation}
where $F_{P}^{inner} = -6 \pi \mu a \mathbf{V}_{P}$, $Z_1 = 6 \alpha^6+3\alpha^5-10\alpha^3+9\alpha-8$ and $Z_2 =  4\alpha^6-9\alpha^5+10\alpha^3-9\alpha+4$. Eq.~\ref{eqn:TD_unstable_drag_force} may also be obtained from the analysis of compound droplets studied in \citet{Sadhal_eccentric}. However, unlike the simple closed form given above, such a calculation from \citet{Sadhal_eccentric} may be tedious\cite{Rushton1973} since the reported solutions are in series form in bispherical coordinates with infinite number of coefficients to be evaluated.
\end{enumerate}

\begin{figure}[]
\centering
  \subfigure[]{
   \includegraphics[height=6.5cm]{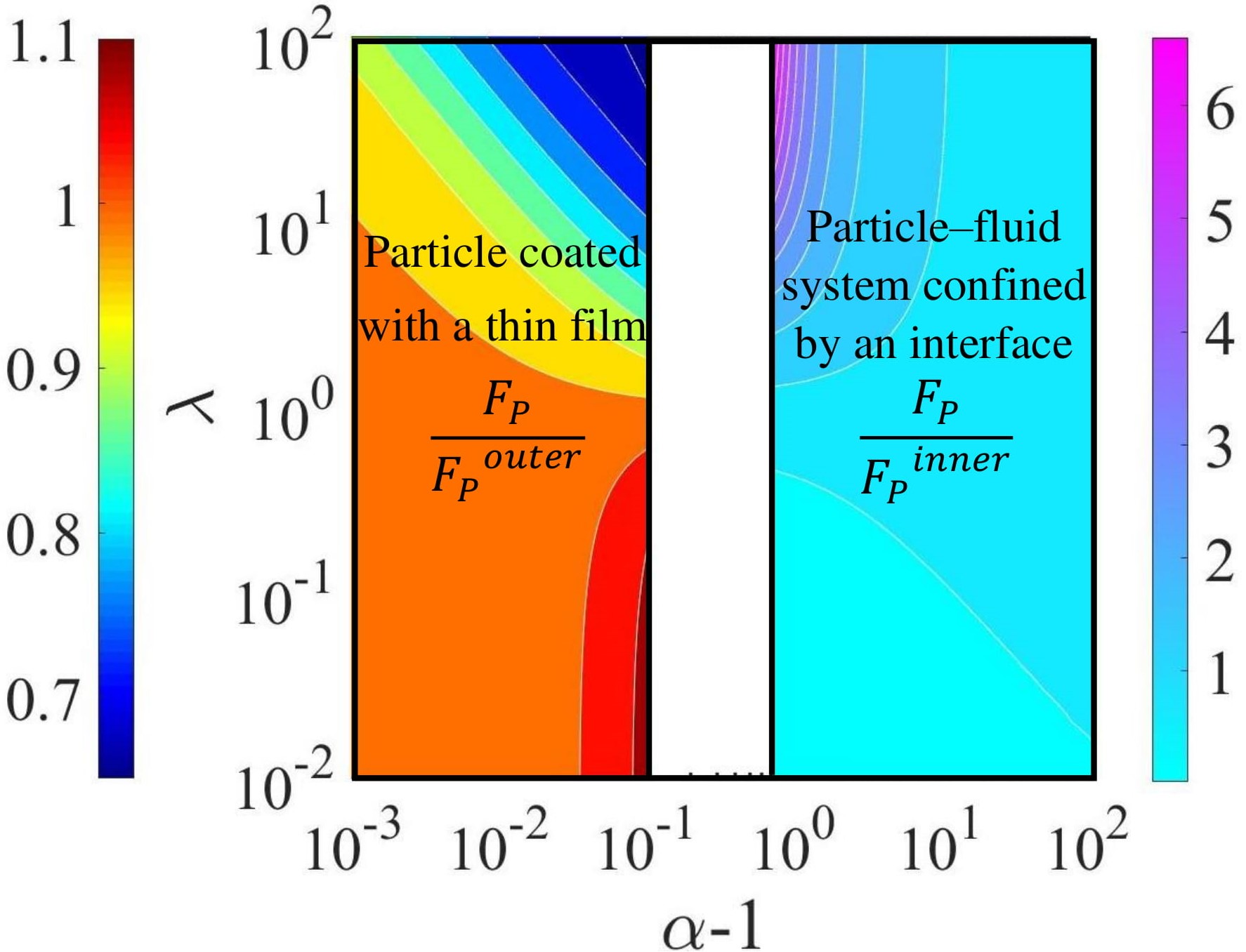}
   \label{fig:TD_drag_contour_1}}
 \subfigure[]{
  \includegraphics[height=6.5cm]{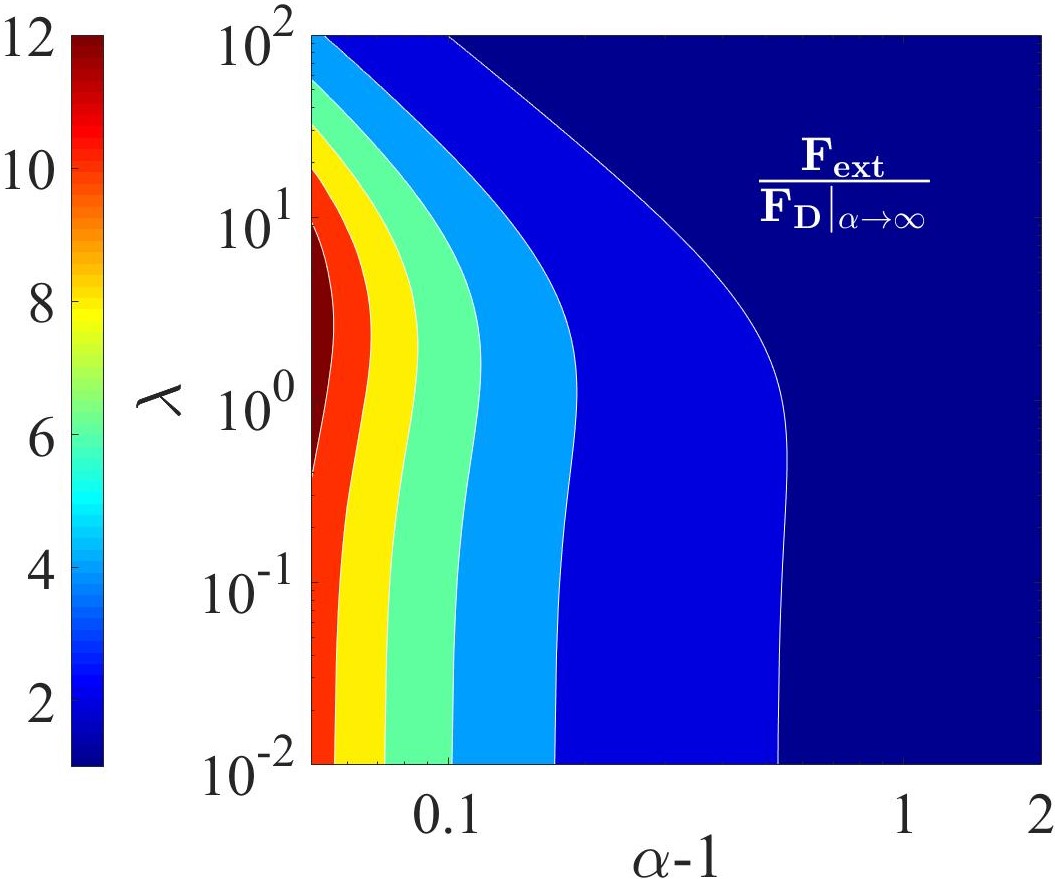}
    \label{fig:TD_stable_drag_force}}
   \caption{(a) (b) Drag force on the particle translating inside a drop as a function of the thickness of the confining drop ($\alpha-1$) and viscosity ratio. Similar to Fig.~\ref{fig:RD_torque} the contour plot on the left hand side shows the modified Stokes drag on a particle due to a thin film coating. The plot on the right hand side shows the modified Stokes drag on a particle when particle-fluid system is confined by an interface. Drag forces are appropriately normalized in each case. (b) Extra force, $F_{ext}$ to be imposed on the inner fluid to make the concentric configuration of a translating compound particle stable as a function of the confining fluid thickness $\alpha - 1$ and viscosity ratio $\lambda$.}
\end{figure}

Modification to the drag force in these two cases is illustrated in Fig.~\ref{fig:TD_drag_contour_1}. Similar to that in Fig.~\ref{fig:RD_torque}, the reduction or increase in the drag force and the contrast in the dependence of viscosity ratio on the modification to the drag force for small and large $\lambda$ can be seen here too.

The translational velocity of the confining drop compared to that of the particle may be obtained as 
\begin{equation}
    \frac{{V}_{D}}{{V_{P}}} = \frac{4+6\alpha^5+\lambda(9\alpha^5-5\alpha^5-4)}{4+6\alpha^5+\lambda Z_1+\lambda^2 Z_2} \quad < 1 .
    \label{TD-drop-vel}
\end{equation}
Therefore, irrespective of the value of $\lambda$ and $\alpha$, the drop always moves with a velocity smaller than that of the particle. Hence the concentric configuration of a translating compound particle cannot be sustained, thus this configuration is unstable. 

\noindent\textbf{Translation of a compound particle with an extra force:} The concentric configuration of a translating compound particle can be made stable. Since the cause for unstable configuration is the slower moving drop, an external body force may be applied on the drop to stabilize the configuration. Therefore, by equating particle and drop velocities, $\mathbf{V}_P = \mathbf{V}_D$ we calculate the extra force required to stabilize the configuration as, $F_{ext} = F_{D} - F_{P} = -8 \pi \mu( \lambda g^{(2)} - g^{(1)})$ and the magnitude of this force is plotted in Fig.~\ref{fig:TD_stable_drag_force}. It may be observed that as the particle becomes much smaller than the drop, the effect of particle on the interface dynamics decreases and therefore the extra force required to stabilize the configuration is simply given by the drag on a drop translating $\left(F_{D}|_{\alpha \to \infty}\right)$ in the outer fluid. Similarly, for a given size ratio, as the viscosity of the inner fluid increases, the drag force on the particle decreases and hence the magnitude of the extra force required to stabilize the configuration decreases. 

The typical velocity field corresponding to the stable concentric configuration of a translating compound particle is shown in Fig.~\ref{fig:TD_stable_vel_contour_100}. It differs from the velocity fields in Fig.~\ref{fig:TD_vel_contour_0}-\ref{fig:TD_vel_contour_100} in two aspects:  (\romannumeral 1)  the velocity field in the inner fluid has a magnitude larger than the two bounding surfaces, i.e., the surface of the particle and the interface, (\romannumeral 2) the fluid inside the drop does not recirculate even for large $\lambda$, unlike that in Fig.~\ref{fig:TD_vel_contour_100}. 

For this stable configuration, the drag force on the compound particle can be calculated as,
\begin{equation}
\frac{-\mathbf{F}_{D}}{4 \pi \lambda \mu \alpha a \mathbf{V}_{P}} =  \frac{6\alpha^3+9\alpha^2+9\alpha+6+\lambda(4\alpha^3+3\alpha^2-3\alpha-4)}{4\alpha^3+6\alpha^2+6\alpha+4+\lambda(4\alpha^3+3\alpha^2-3\alpha-4)}.
    \label{TD-stable-dragforce}
\end{equation}
Eq.~\ref{TD-stable-dragforce} can also be derived from \citet{Rushton} by taking the limit of infinite viscosity for the fluid inclusion of a translating compound droplet.
Again, it is useful to look at the limit of $\alpha \to 1$, the case of a rigid particle  coated with a thin fluid film. Then the above expression reduces to,
\begin{equation}
    \mathbf{F}_{D} = -6 \pi \lambda \mu a \mathbf{V}_P \left(1+(\alpha-1) \Big[1-\frac{1}{4}\lambda\Big]\right)
    \label{TD_stable_drag_force_thin}
\end{equation}
which is in agreement with that given by Choudhuri\cite{Choudhuri2010}. Interestingly, for $\lambda > \frac{1}{4}$ the drag force on a coated sphere is smaller than that on the sphere itself. Johnson \cite{johnson_1981} also had concluded the same in a similar analysis. Note that in the absence of a stabilizing external force, Eq.~\ref{eqn:TD_unstable_thin_film_drag_force}  predicts a reduction in the drag force on the particle only when $\lambda > 1$. 

\subsection{A compound particle in a general linear flow}
\label{sec:linearflow}

\begin{figure}[]
 \centering
 \subfigure[]{
  \includegraphics[height=4cm]{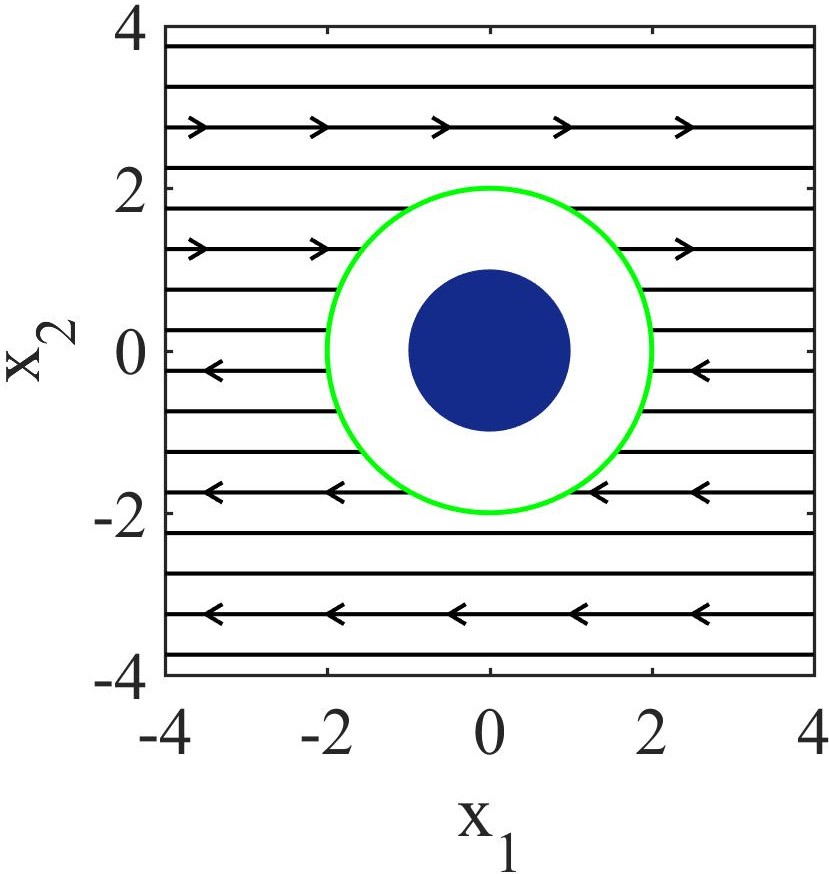}
  \label{fig:shear_schematic}}
   \subfigure[]{  \includegraphics[height=4cm]{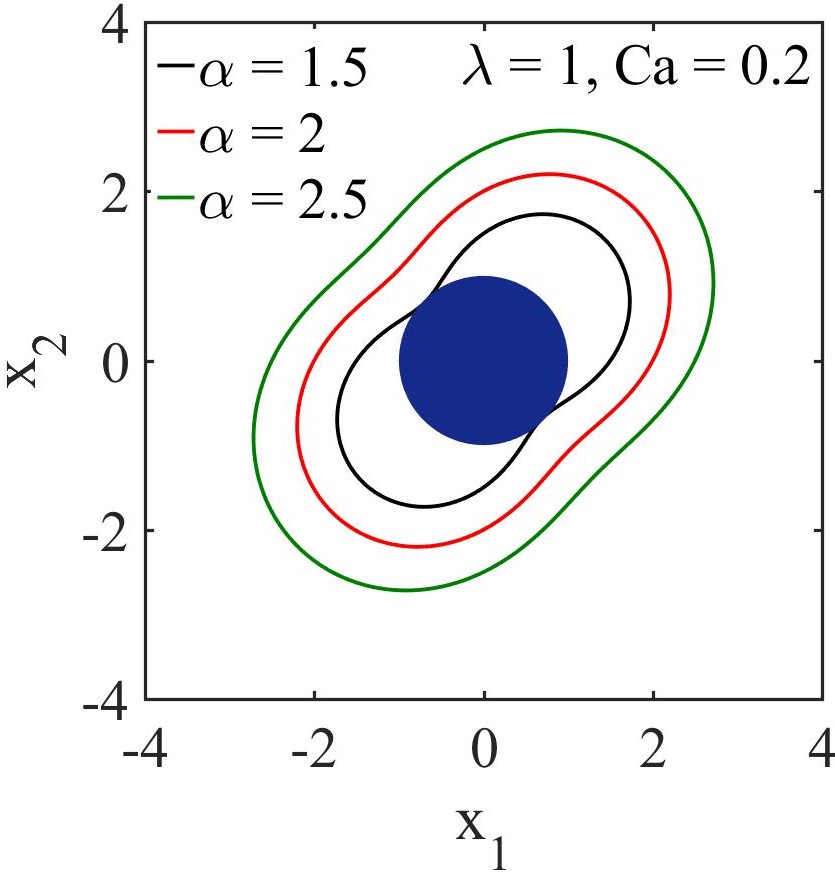}
\label{fig:ss_drop_shapes_1}}
   \subfigure[]{
  \includegraphics[height=4cm]{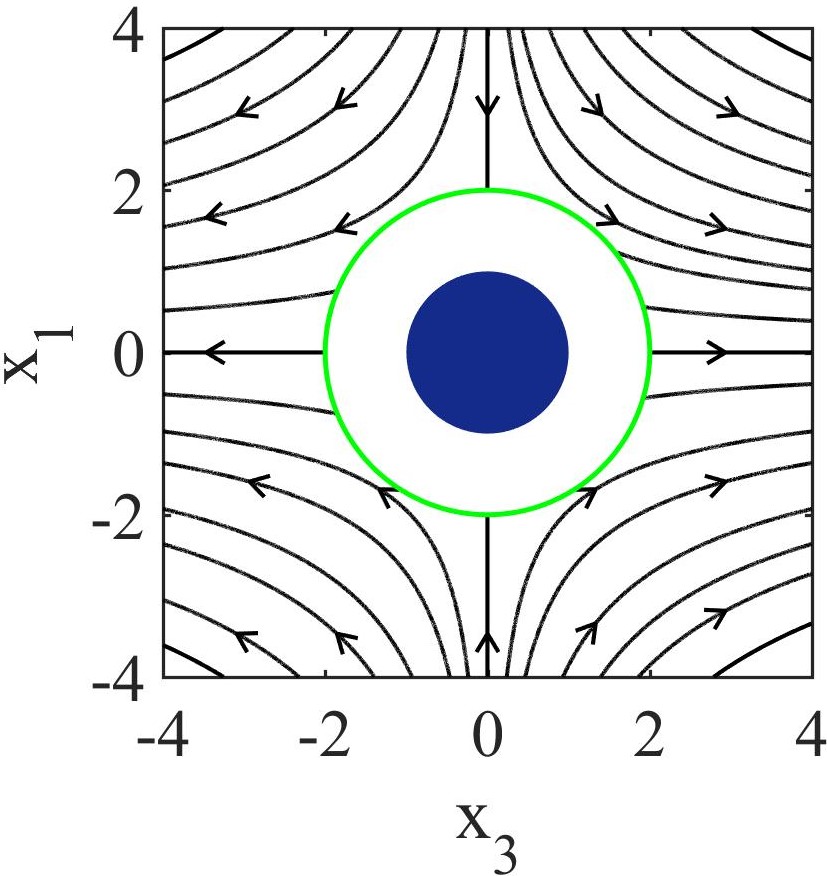}
   \label{fig:uniaxial__schematic}} 
        \subfigure[]{
  \includegraphics[height=4cm]{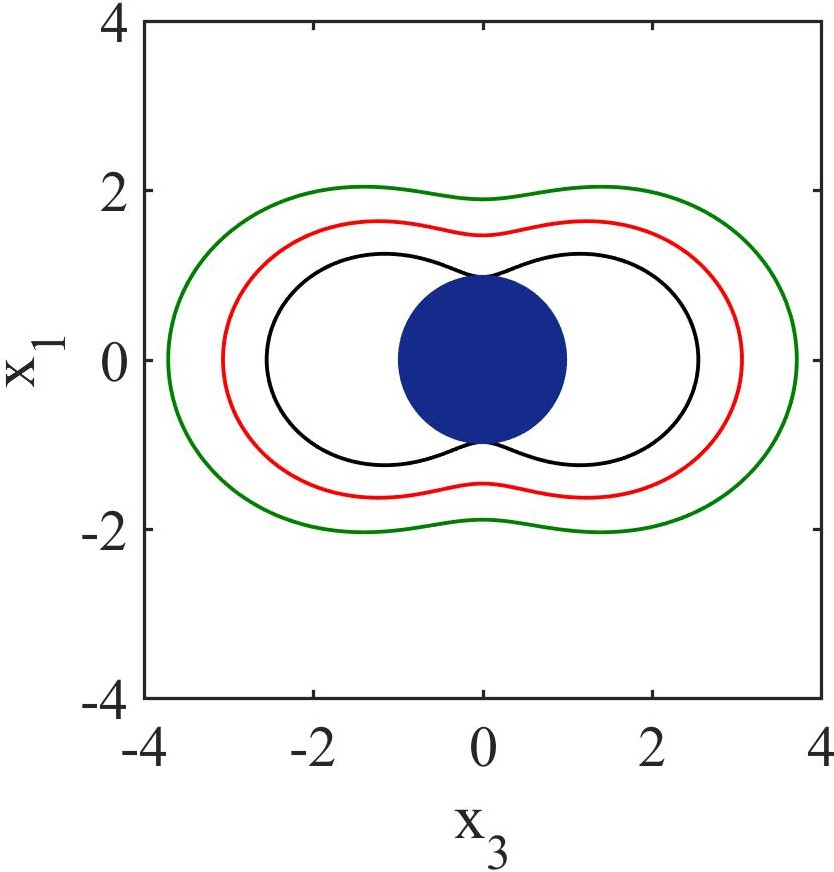}}
     \subfigure[]{
   \includegraphics[height=4cm]{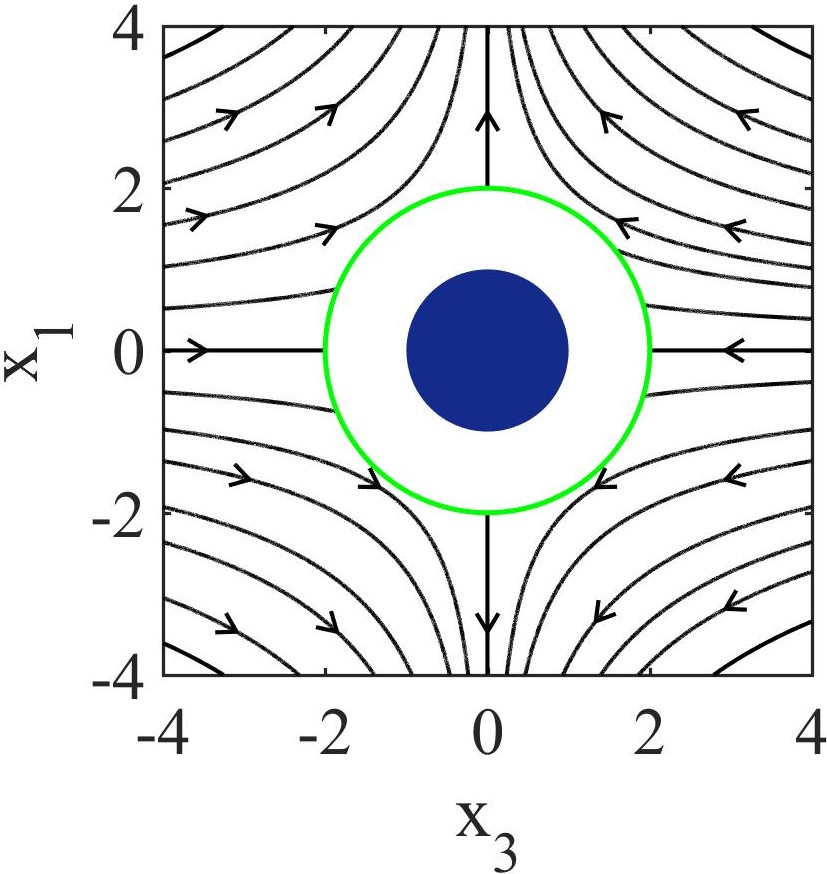}
   \label{fig:biaxial__schematic}}
  \subfigure[]{
  \includegraphics[height=4cm]{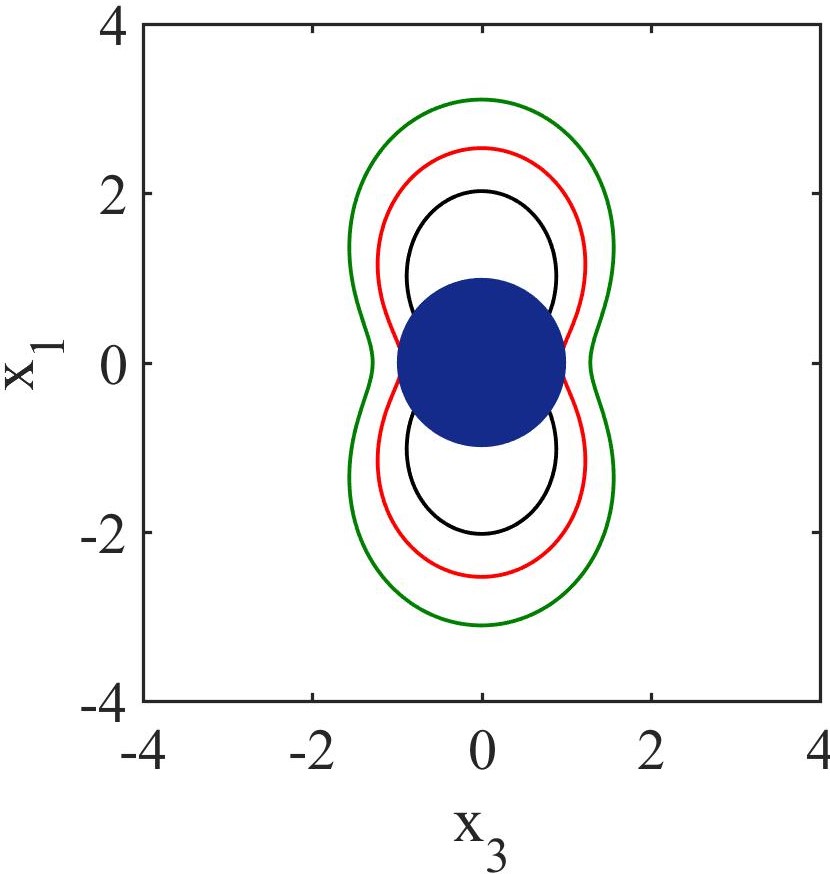}}
 
    \caption {Typical linear flow fields encountered in a microfluidic system and the resulting steady state shapes of the compound particle in these flows. In a right handed coordinate system ($x_1, x_2, x_3$) (a) simple shear flow, obtained as $\mathbf{u}^{imp} = \mathbf{x} \cdot \nabla \mathbf{u}^{imp} = \mathbf{x}\cdot(\mathbf{E}+\boldsymbol{\Omega})$ where $E_{12} = 1/2, \ E_{21} = 1/2$, $\Omega_{12} = 1/2, \ \Omega_{21} = -1/2$ and other elements of $E_{ij}, \ \Omega_{ij}$ being zero. (c) Uniaxial flow is obtained by having $E_{11} = -1/2, \ E_{22} = -1/2, \ E_{33} = 1$ and other elements to be zero. (e) Biaxial flow is obtained by reversing the uniaxial flow, $E_{11}  = 1/2, \ E_{22} = 1/2, \ E_{33} = -1$ and other elements are zero. (b),(d) and (f) Corresponding drop shapes, drawn for $\lambda =1$, $Ca = 0.2$ and for various $\alpha$. For $\alpha = 1.5$ breakup occurs  in all flows but for $\alpha = 2$ it occurs  only in biaxial flow.}
 \label{fig:imposedflows}
 \end{figure}

In order to understand the behaviour of compound particles in imposed flows, we will select some typical linear flow fields encountered in microfluidic devices. This is achieved by adopting various forms for $\nabla \mathbf{u}^{imp}$ as shown in Fig.~\ref{fig:imposedflows}. Simple shear flow is a good approximation to flow near rigid surfaces, both uniaxial and biaxial extensional flows are typically encountered at various junctions (e.g., Y, T) in a microfluidic device.

\begin{figure*}[t]
\centering
  \subfigure[]{
  \includegraphics[height=5.0cm]{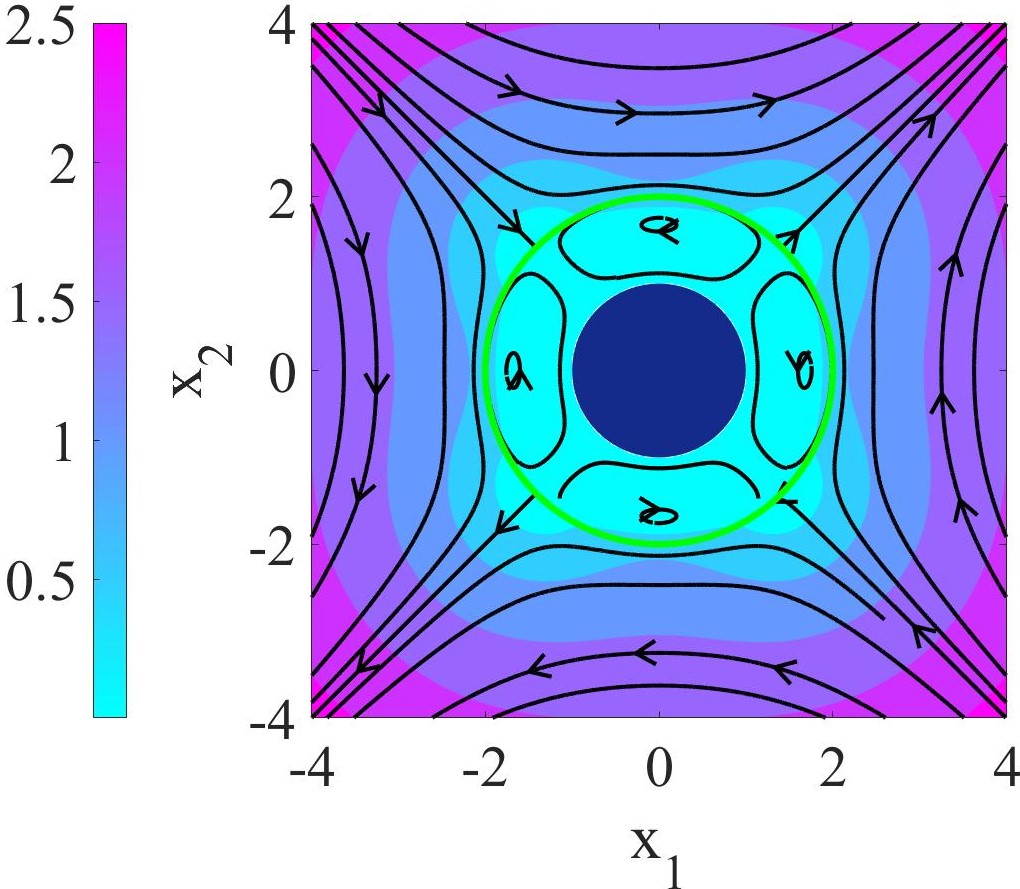}
      \label{fig:ss_flow_field}}\quad
\subfigure[]{
  \includegraphics[height=5.0cm]{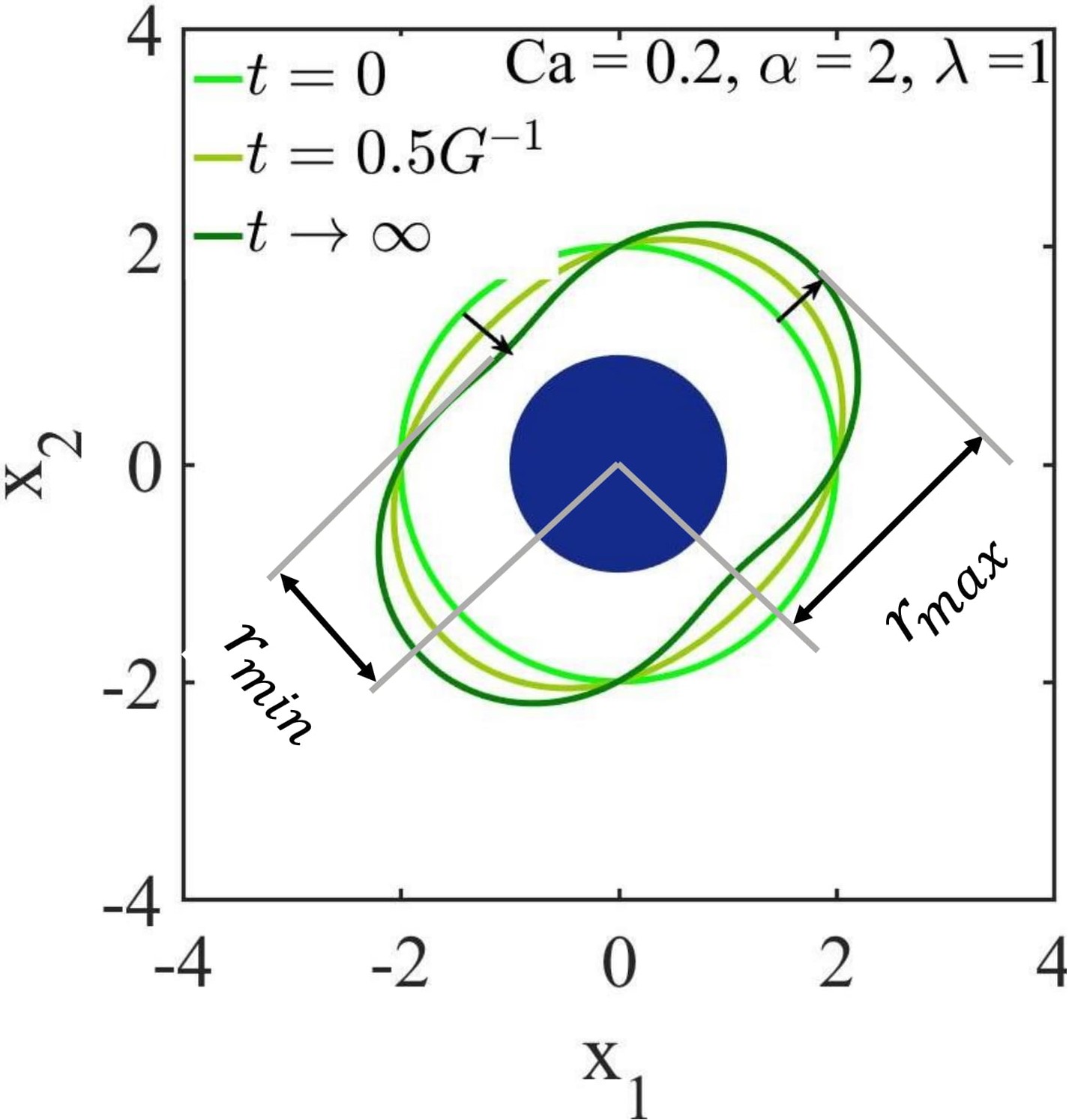}
  \label{fig:ss_drop_evolution}}\quad 
\subfigure[]{
  \includegraphics[height=5.0cm]{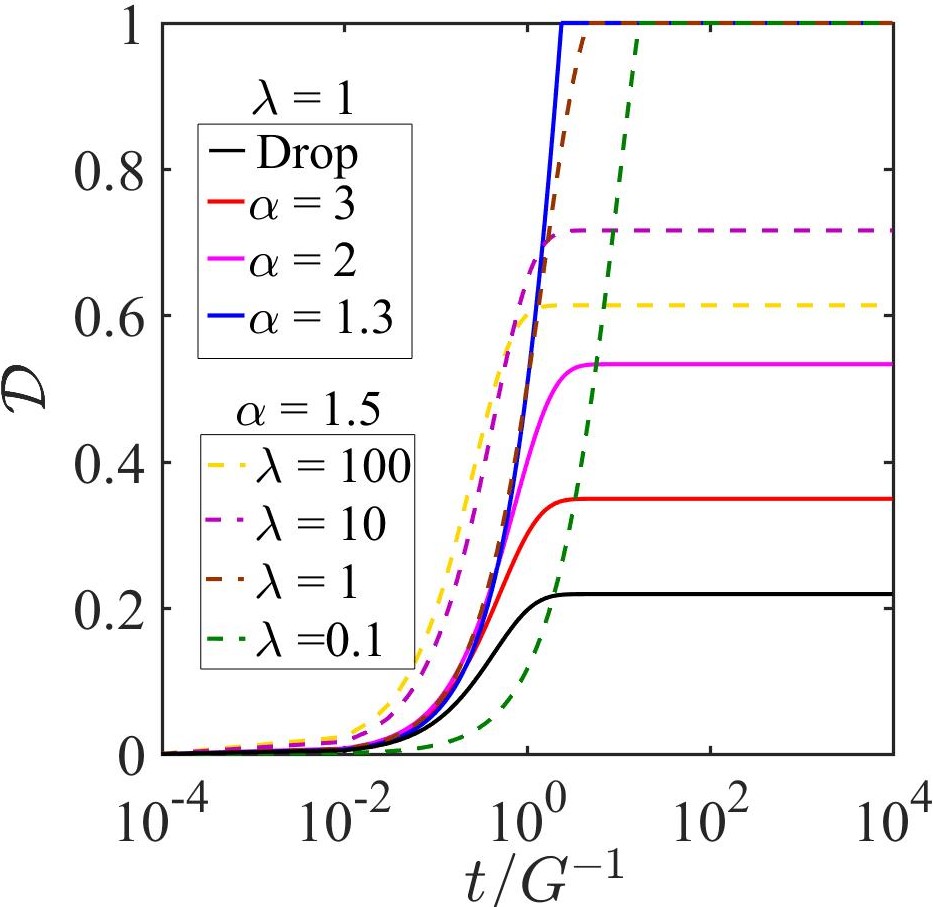}
  \label{fig:ss_dvst_n}}\\
\subfigure[]{
  \includegraphics[height=5.0cm]{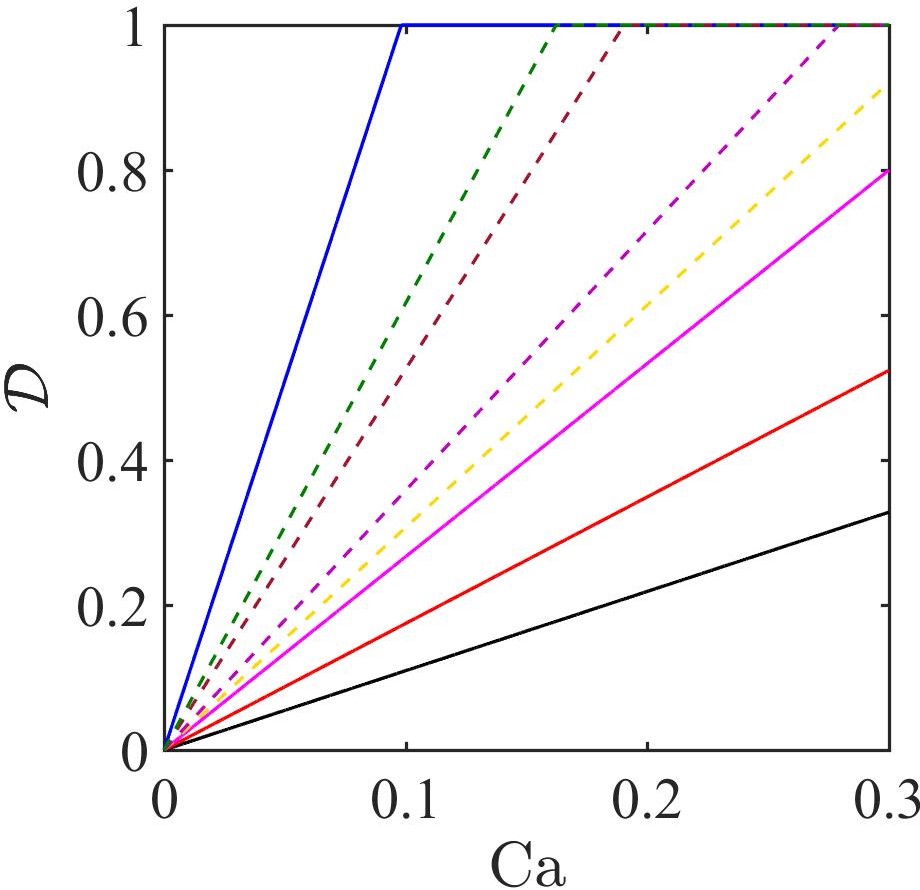}
  \label{fig:ss_dvsca_n}}\quad
  \subfigure[]{
  \includegraphics[height=5.0cm]{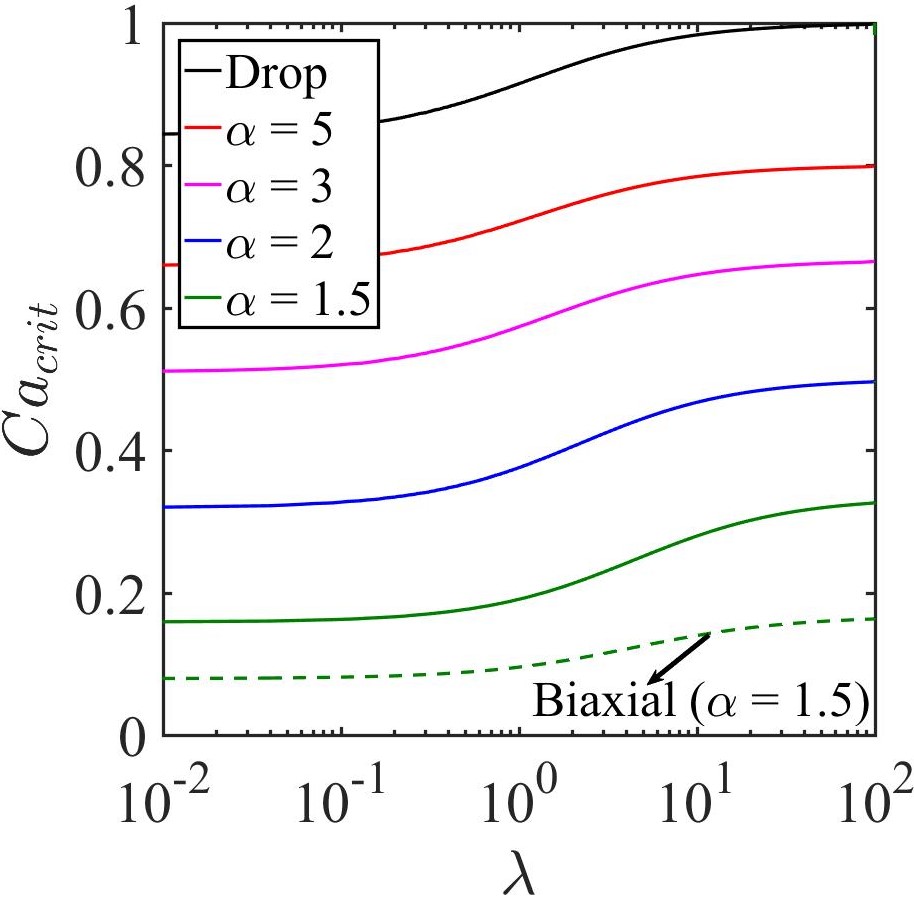}
    \label{fig:ss_phase_plane}}\quad
\subfigure[]{
  \includegraphics[height=5.0cm]{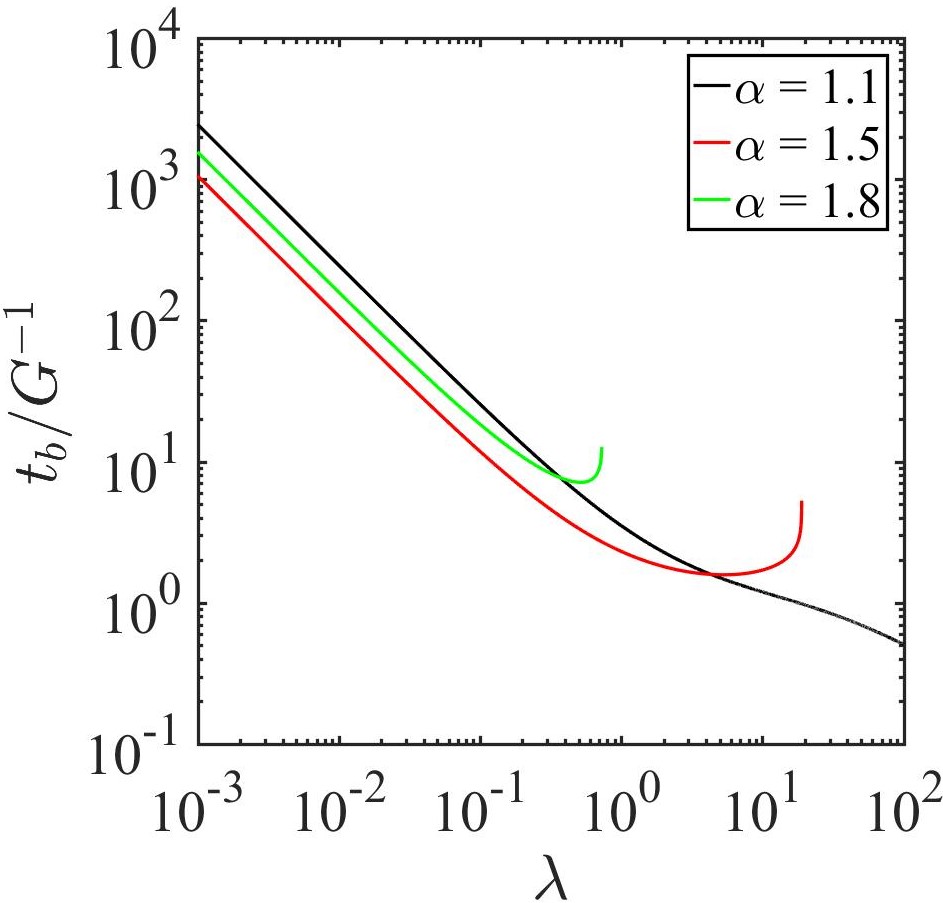}
  \label{fig:ss_tbvslam}}
   \caption{(a) Velocity field around a compound particle in an imposed simple shear flow in $x_1 - x_2$ plane, for $\alpha = 2,\  \lambda = 1,$ and $Ca = 0.2$. The domain is colored according to the magnitude of velocity and streamlines are superimposed onto it. Response to only the symmetric part of the velocity gradient tensor $\mathbf{E}$ is shown. (b) The confining drop deforms from a spherical shape to an elongated shape in a simple shear flow. (c) Time evolution of deformation parameter, $\mathcal{D}$ at $Ca = 0.2$ for various $\alpha$ and $\lambda$. Time is normalized with the inverse shear rate, $G^{-1}$. (d) Steady state values of deformation parameter as a function of $Ca$ for various $\alpha$ and $\lambda$, legend same as that in (c). (e) $Ca_{crit}$ as a function of $\lambda$ for various values of $\alpha$. (f) Time required for the breakup $t_{b}$ of the confining drop as a function of $\lambda$ for various $\alpha$ at $Ca = 0.3$.} 
\end{figure*}

We will first discuss our results in the context of simple shear flow and then compare them with those of both uniaxial and biaxial flows.
A compound droplet subjected to a linear flow has been studied before \cite{Brenner,lealstone} but these works ignore the temporal deformation dynamics of the confining interface. As we see below, understanding temporal dynamics is important as it paves the way to stabilize the configuration of a compound particle, which is explored in this work.
Thus, as the third problem, we consider a compound particle subjected to a linear flow, $\mathbf{E} \cdot \mathbf{x}$ where $\mathbf{E}$ is the symmetric part of the imposed velocity gradient tensor, $\nabla \mathbf{u}^{imp}$. The corresponding anti-symmetric part (say, $\boldsymbol{\Omega}$) gives rise to only a solid body rotation of the compound particle and it will not have any other dynamical consequences. The velocity and pressure fields may be obtained as,

\begin{fleqn}[0pt]
\begin{align*}
\mathbf{u}^{(1)} &= \Bigg(d_1-6r^2d_2-6\frac{c_{3}^{(1)}}{r^5}\Bigg)\mathbf{E}.\mathbf{x} +\Bigg(\frac{12}{5}d_2+\frac{c_{1}^{(1)}}{2r^5} \\+15\frac{c_{3}^{(1)}}{r^7}\Bigg)\mathbf{x}.(\mathbf{x}.  \mathbf{E}.\mathbf{x})\nonumber,  \hspace{-3cm} 
\end{align*}
\end{fleqn}

\begin{align}
\mathbf{u}^{(2)} &= \Bigg(1-6\frac{c_3^{(2)}}{r^5}\Bigg)\mathbf{E}.\mathbf{x}+\Bigg(\frac{c_{1}^{(2)}}{2r^5}+15\frac{c_{3}^{(2)}}{r^7}\Bigg)\mathbf{x}(\mathbf{x}.\mathbf{E}.\mathbf{x}) \nonumber,\\ 
 p^{(1)} &= c_{1}^{(1)} \mathbf{E} \ : \ \Big(\frac{\mathbf{xx}}{r^5}\Big)+d_2
 \mathbf{E}:\Big(\mathbf{xx}\Big), \quad 
 p^{(2)} = c_{1}^{(2)} \mathbf{E} \ : \ \Big(\frac{\mathbf{xx}}{r^5}\Big).\nonumber
\end{align}
\noindent The expressions for the constants $c_{i}^{(k)}$ and $d_{i}$ are given in appendix B. For a compound particle in an imposed simple shear flow, the velocity field obtained is shown in Fig.~\ref{fig:ss_flow_field}.

\noindent\textbf{Evolution of confining drop shape:} Unlike the solution of rotating and translating compound particle where the spherical shape of the confining surface was an exact solution, here the drop is deformed due to the extensional (or compressional) nature of the flow field.
 Defining \cite{leal2007advanced} $S(\mathbf{x},t)  = r- \alpha a (1 + b~Ca~\mathbf{x} \cdot \mathbf{E} \cdot \mathbf{x} )$ to describe the interface, kinematic boundary condition (Eq.~\ref{LF-kinematic-condition}) provides the evolution of the interface in terms of a single scalar parameter $b$,
\begin{equation}
    Ca \frac{\partial b} {\partial t} = - \frac{b g(\alpha,\lambda)-f(\alpha,\lambda)}{h(\alpha,\lambda)},
    \label{LF-Ca-kinematic-condition}
\end{equation}
where $f(\alpha,\lambda) = 5\lambda[2(19+16\lambda)\alpha^{10}-25(8\lambda-1) \alpha^7 +\alpha^5(336 \lambda-231)-200(\lambda-1)\alpha^3+32(\lambda-1)]$,  $g(\alpha,\lambda) = 4[4\alpha^{10}(\lambda+1)-5\alpha^7(5\lambda+2)+42\alpha^5\lambda-5\alpha^3(5\lambda-2)+4(\lambda-1)]
$, and 
$h(\alpha,\lambda)  = 2[\alpha^{10}(48\lambda^2+89\lambda+38)-75\alpha^7(4\lambda^2-\lambda-3)+168\alpha^5(3\lambda^2-\lambda-2)-100\alpha^3(3\lambda^2-\lambda-2)+48(\lambda-1)^2].$ Therefore,  interface evolution is an  exponential relaxation over a time scale of $\tau = Ca~{h(\alpha,\lambda)}/{g(\alpha,\lambda)}$
in response to the generated fluid flow. In the limit of $\alpha \to 1$, $\tau$ is given by, 
\begin{equation}
    \tau = \frac{Ca}{8\lambda (\alpha-1)^3}+O\Big(\frac{1}{(\alpha-1)^2}\Big).
    \label{tau_film}
\end{equation}
Therefore, as $\lambda \to 0$ or $\alpha \to 1$, the fluid film coated on the surface of the particle strongly resists deformation (film behaves like the enclosed rigid particle).

The evolution of the confining drop shape in response to a simple shear flow is illustrated in Fig.~\ref{fig:ss_drop_evolution}. The extensional part of the flow elongates the drop as shown in the figure. Since the fluids are incompressible, the confining drop shrinks in the orthogonal directions. Thus a spherical drop deforms to a shape similar to a dumb-bell. Following Taylor \cite{Taylor}, it is convenient to define a deformation parameter in order to describe the deformed drop shape,
\begin{equation}
    \mathcal{D} = \frac{(r_{max}-a)-(r_{min}-a)}{(r_{max}-a)+(r_{min}-a)}.
  \label{SS_deformation_parameter_equation}
\end{equation}
Here $r_{max}$ and $r_{min}$ respectively represent the longest and shortest lengths of the confining drop. $\mathcal{D}=0$ corresponds to the spherical confining drop and $\mathcal{D}=1$ corresponds to the case when the drop touches the solid inclusion. 

Both the evolution time scale and the final shape are dependent on  the capillary number, $Ca$, the size ratio of the drop to the particle, $\alpha$, and the viscosity ratio of the outer fluid to the inner fluid, $\lambda$. Fig.~\ref{fig:ss_dvst_n} shows the temporal evolution of deformation parameter for various values of $\alpha$ and $\lambda$. 
When $\alpha$ is large, the confining drop is much bigger than the solid inclusion and $\mathcal{D}$ is small. As $\alpha$ decreases, $\mathcal{D}$ increases indicating larger deformation of the confining drop. It is possible that the deformed drop comes into contact with the solid particle ($\mathcal{D} = 1$) which may be taken as an indication of breakup of the confining drop. A similar behavior of increased deformation is observed with decrease in $\lambda$ also.  Therefore, everything else remaining the same, a confining drop of larger viscosity will deform more than a drop of smaller viscosity. This non-typical behaviour arises from the fact that increasing the viscosity of the inner fluid increases the viscous resistance to flow and the consequent larger viscous stress at the interface results in larger deformation of the drop. 
As expected, increasing the strength of imposed flow (increased $\mathbf{E}$) also increases the extent of deformation. This is illustrated in Fig.~\ref{fig:ss_dvsca_n} where $\mathcal{D}$ corresponding to the steady state is plotted as a function of $Ca$ for various $\alpha$ and $\lambda$.

\begin{table*}
\begin{center}
\renewcommand{\arraystretch}{3}     
\small\addtolength{\tabcolsep}{-0pt}
\begin{tabular}{|c|c|c|c|c|c|}
\hline
& $r_{int}$ & $r_{max}$ & $r_{min}$ & $\mathcal{D}$ & $ t_b/G^{-1}$ \\ \hline
Simple shear & $a\alpha\left(1+b \ Ca \ \sin^{2}\theta \sin \phi cos \phi\right)$ & $a \alpha \left(1+\dfrac{1}{2}b \ Ca\right)$ & $a \alpha \left(1-\dfrac{1}{2}b \ Ca\right)$ &$\dfrac{b \ Ca}{2-2\alpha}$ &$ -\tau \ \log\Big(1-\dfrac{2(1-\alpha)}{\alpha \ b \ Ca}\Big)$ \\  \hline
Uniaxial &$a\alpha \Bigg(1+b \ Ca  \left(\dfrac{3}{2}\cos^2 \theta-\dfrac{1}{2}\right)\Bigg)$ & $ a \alpha \left(1+b \ Ca\right)$ &$ a \alpha \left(1-\dfrac{1}{2}b \ Ca\right)$ &$ \dfrac{1.5 \  b \ Ca}{2+0.5 \ b \ Ca-2\alpha}$ &$ -\tau \  \log\Big(1-\dfrac{2(1-\alpha)}{\alpha \ b \ Ca}\Big)$ \\ \hline
Biaxial &$ a\alpha \Bigg(1-b \ Ca \left(\dfrac{3}{2}\cos^2 \theta-\frac{1}{2}\right)\Bigg)$&$ a \alpha \left(1-\dfrac{1}{2}b Ca\right)$&$a \alpha \left(1-b \  Ca\right)$& $ \dfrac{1.5 \ b  \ Ca}{2-0.5  \ b  \ Ca-2 \ \alpha}$&$-\tau  \ \log\Big(1-\dfrac{(1-\alpha)}{\alpha \ b \ Ca}\Big)$ \\ \hline
\end{tabular}
\end{center}
\caption{Comparison of shape of the deforming drop in terms of location of the interface $r_{int}$, $r_{max}$, $r_{min}$, the extent of deformation $\mathcal{D}$ and the breakup time $t_b$. Here, $\theta$ and $\phi$ respectively represent the polar and azimuthal angles defined in the spherical coordinates.}
\label{tab:table1}
\end{table*}

\noindent\textbf{Breakup of the confining drop:} The stronger deformations can result in a breakup of the confining drop. This can be noted when the interface comes into contact with the solid inclusion. As seen in Fig.~\ref{fig:ss_dvsca_n} there exists a critical capillary number $Ca_{crit}$ beyond which $\mathcal{D} = 1$ for a given value of $\alpha$ and $\lambda$. From Eq.~\ref{SS_deformation_parameter_equation}, the $Ca_{crit}$ in the limit of $\alpha \to 1$ can be calculated (for $\lambda~\sim O(1)$) as,
\begin{equation}
    Ca_{crit} = \frac{32}{15} (\alpha-1)^2+O(\alpha-1)^3.
    \label{crit_ca_film}
\end{equation}
Therefore, in the case of very thin films, the strength of flow required for breakup is proportional to the square of the thickness of the confining drop. The complete dependence of $Ca_{crit}$ on the size ratio $\alpha$ and the viscosity ratio $\lambda$ is illustrated in Fig.~\ref{fig:ss_phase_plane}. If the viscosity of the inner fluid is much larger than that of the outer fluid, $Ca_{crit}$ is almost independent of $\lambda$. In the other limit, when the viscosity of the inner fluid is much smaller than that of the outer fluid, $Ca_{crit}$ increases monotonically with decrease (increase) in the viscosity of the inner (outer) fluid. Finally when $\lambda \to \infty$ the viscosity of the inner fluid is very small so that the additional stress due to the presence of the solid in the inner fluid is not felt by the interface and $Ca_{crit}$ becomes independent of $\lambda$ again. Certainly, increasing the size of the confining drop increases $Ca_{crit}$ required for breakup. Therefore, a stable configuration of the compound particle exists only when $Ca < Ca_{crit}$ while $Ca_{crit}$ itself can be increased by increasing the size and viscosity ratio. For example, it is always possible to have a stable thin film coated on a particle when the film is a fluid of low viscosity.

The viscosity of the inner and the outer fluids play an important role in determining the rate of deformation of the confining drop as well. If the viscosity of the inner fluid is very large ($\lambda \to 0$) then the dynamics is primarily determined by the viscosity of the inner fluid and deformation process of the interface is very slow due to damped dynamics. On the other hand, if the inner fluid is less viscous compared to the outer fluid ($\lambda \to \infty$) the confining drop dynamics and the deformation process is relatively faster. This difference in the rate of deformation for various $\lambda$ is seen in Fig.~\ref{fig:ss_dvst_n}. We will now quantify this time scale by defining the breakup time $t_b$.

The time required for the breakup of the confining drop, $t_b$ can be determined from Eq.~\ref{LF-Ca-kinematic-condition} and Eq.~\ref{SS_deformation_parameter_equation}  and the dependence of $t_{b}$ on $\alpha$ and $\lambda$ is shown in Fig.~\ref{fig:ss_tbvslam}. As the viscosity of the inner fluid decreases, $t_{b}$ decreases due to reduction in the viscous resistance to the flow. However beyond a particular viscosity ratio ($\lambda \sim 1$), $t_{b}$ increases with increase in $\lambda$. This is because, large viscosity of the outside fluid slows down the confining interface dynamics. Comparison of $t_b$ for various $\alpha$ also shows a non-monotonic trend with $\alpha$. In the limit $\alpha \to 1$, $t_b$ is given as,
\begin{equation}
\frac{t_b}{G^{-1}} = -\frac{4}{15 \lambda (\alpha-1)}-\frac{4(15\lambda-7)}{45\lambda}+O(\alpha-1)
\label{tb_film}
\end{equation}
suggesting that the time taken for the thin film to breakup is very large. This occurs because, the rigid particle suppresses the fluid flow in the thin film. However as $\alpha$ increases, breakup time again increases due to larger film thickness.
 
\noindent\textbf{Drop deformation with and without a solid inclusion:} It will be interesting to compare the deformation dynamics of a compound particle with that of a drop (without a solid particle). This comparison allows us to isolate the role of solid inclusion in drop dynamics. Both the rate of deformation and the steady state shape are affected by the presence of solid particle. As seen in Fig.~\ref{fig:ss_dvst_n}, rate of deformation is least for a drop and it increases with increase in $\alpha$. Similarly, the deformation parameter $\mathcal{D}$ shown in Fig.~\ref{fig:ss_dvsca_n} is also the least for a drop at a given $Ca$. In other words, $Ca_{crit}$ is highest for a drop (Fig.~\ref{fig:ss_phase_plane}). Physically this can be understood as follows. Keeping the drop radius same, increase in the size of the particle increases the viscous resistance to the flow. Consequently, for the same imposed flow, the stress at the interface is larger resulting in a larger deformation. Secondly, the breakup of the confining drop is marked when the drop interface comes into contact with the solid inclusion. In the absence of the particle, sufficiently high deformation of the interface is required so that two sides of the drop meet each other resulting in a breakup. On the other hand, for a compound particle, breakup occurs when the deformed interface meets the solid surface. This can be achieved with much weaker deformation of the interface. In other words, presence of a solid particle inside a drop makes it susceptible to breakup.

A clarification regarding the smallness of thickness of the film in case of thin film limits analyzed is in order here. 
In the limit $\alpha \to 1$, we can calculate $r_{max}$ as,
\begin{equation}
\frac{r_{max}}{a} = \frac{15}{32}\Big(\frac{Ca}{\alpha-1}\Big)+O(\alpha-1).
\end{equation}
Therefore, when $\alpha \to 1$, $r_{max}/a \sim O(1)$ if $(\alpha-1) \sim O(Ca)$. In other words, the thickness of the thin films have to be $O(Ca)$ for the expressions to hold when $\alpha \to 1$ limit is applied. Otherwise the analysis holds for all $\alpha$.

\subsection{Comparison between imposed flows} 

So far we have analyzed the response of the compound particle in a simple shear flow. We now make a comparison between simple shear, uniaxial and biaxial flows. The results are summarized in Fig.~\ref{fig:imposedflows} by plotting the drop shapes and in Table \ref{tab:table1} in terms of expressions for the confining drop shape ($r_{int}$), the longest ($r_{max}$) and the shortest ($r_{min}$) dimensions of the confining drop, the deformation parameter ($\mathcal{D}$) and the breakup time ($t_b$). The qualitative dependencies of $\alpha$ and $\lambda$ on the drop deformations are same for all flows. Neither the onset of breakup characterized by $r_{min}$ nor the time required for breakup $t_b$ are different for simple shear and uniaxial flows. As expected, uniaxial flows are strongly stretching (largest $r_{max}$) and biaxial flows are strongly compressive (smallest $r_{min}$). Interestingly, the breakup time and the $Ca_{crit}$ for a biaxial flow is the least. This strong effect of biaxial flows in deforming the confining drop and the resulting breakup can also be observed in Fig.~\ref{fig:imposedflows}. This effect may be understood as follows. Biaxial flow compresses the drop towards an oblate shape while uniaxial flow stretches the drop towards a prolate shape. The oblate shaped interface can meet the solid inclusion faster than the corresponding prolate. Therefore, though biaxial flow is obtained by reversing the direction of uniaxial flow, the former is stronger than the latter for the breakup of confining drop.

\subsection{Pulsatile flows to prevent breakup of confining interface}

Since we have seen that a confining drop in a compound particle is susceptible to breakup, it will be interesting to investigate the conditions that prevent this. Here, we show that manipulating the time scale of the imposed flow field is a solution to maintain the stability of the compound particle. Since interface evolution occurs over a time scale of $\tau$, imposing a pulsatile flow helps as described below.

\begin{figure}[t]
\centering
\subfigure[]{\includegraphics[height=3.9 cm]{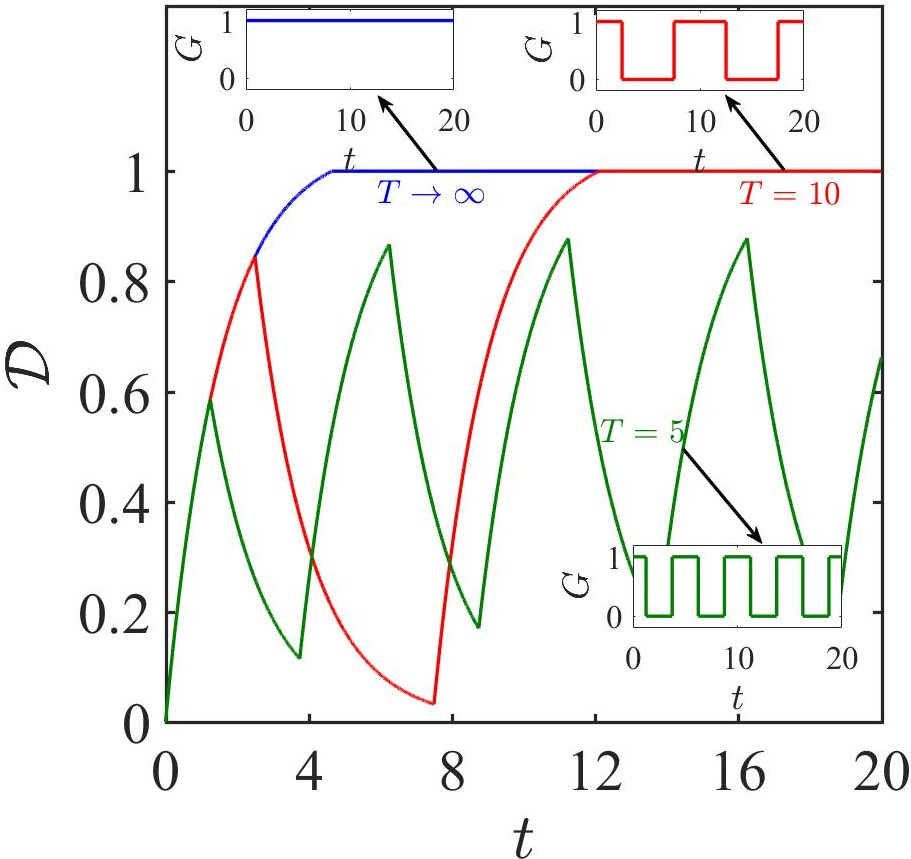}
      \label{fig:squarewaveD}}
\subfigure[]{\includegraphics[height=3.9 cm]{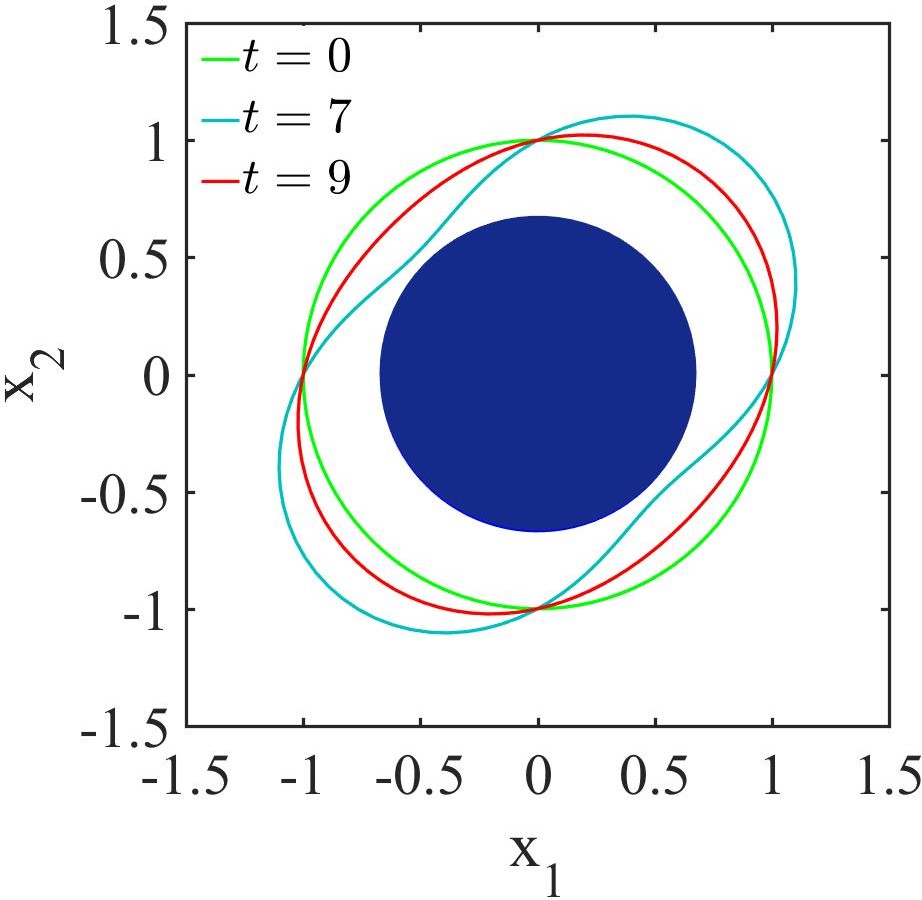}    \label{fig:step_drop_shape}}
   \caption{(a) Deformation dynamics, quantified as $\mathcal{D}$ of the confining drop of a compound particle when subjected to a pulsatile shear flow with time period $T = 5$, $T = 10$ and a non-pulsatile flow ($T \to \infty$). Here $Ca = 0.2$, $\alpha = 1.5$, $\lambda = 1$. The corresponding insets show time variation of strength of the imposed flow $G(t)$. $t$ is normalized by the inverse shear rate in all cases. (b) The shape of the confining drop at three different instances of the square wave input for $T = 5$. }
   \label{fig:pulsatile}
\end{figure}

Consider an imposed, periodic shear flow described by a strain rate $G(t)$, of the square wave form, 
\begin{equation}
    G (t) = \frac{1}{2} \left(1+sgn\left[\cos( \frac{2 \pi t}{T})\right]\right),
\end{equation}
where $T$ is the time period of the wave and $sgn$ is a signum function. Therefore, $G = 1$ and $G = 0$ correspond to, respectively, the presence and absence of the imposed flow. The input forms of the wave and the corresponding drop shapes quantified as $\mathcal{D}(t)$ are plotted in Fig.~\ref{fig:pulsatile}. Drop deformation $\mathcal{D}(t)$ corresponding to a simple shear flow is also given for comparison. 

In the case of simple shear flow, shear field  continuously deforms the interface. The confining drop finally comes into contact with the solid inclusion, represented as $\mathcal{D} = 1$, and it marks the breakup of the confining drop as seen earlier. 

However, the interface takes a time scale of $\tau$ to evolve towards the new shape as illustrated by the exponential curve $\mathcal{D}(t)$. Therefore, if a square wave flow is chosen, the drop deforms during the period in which the fluid flow prevails ($G=1$) but before the interface meets the solid inclusion, flow is turned off ($G = 0$).

 The drop then relaxes to a spherical shape driven by surface tension force. The drop starts deforming again on imposition of the next wave and the process continues. If $T$ is large (as in $T = 10$ in Fig.~\ref{fig:squarewaveD}), the confining drop meets the solid inclusion during the subsequent cycles and the breakup happens. If  $T$ is sufficiently small (as in $T = 5$ in Fig.~\ref{fig:squarewaveD}) then the oscillations continue in every cycle without the drop ever touching the limit $\mathcal{D} = 1$ and the confining drop does not break. While the flow in the system results in transportation of the compound drop, the pulsatile nature of the flow preserves the drop from breakup.
Therefore, operating parameters may be carefully tuned, for e.g., in a microfluidic device, such that $T > \tau$, the drop does not breakup in time $T$ and the stability of the compound particle can be maintained. Of course both $\tau$ and $T$ must be much larger than the inertial time scale $\rho a^2/\mu$ for the above analysis to hold.
\section{Conclusion and outlook}
In this work, we studied the dynamics of a compound particle in a concentric configuration when the particle is acted upon by an external force, torque or a linear flow. The incompressible Stokes equations were solved analytically to obtain the exact solutions for the flow field. The concentric configuration of a rotating compound particle is found to be stable, but that of a translating particle is only transient, as the drop moves slower than the particle. 
The force and torque acting on the compound particle were calculated which further illustrated two interesting cases: 1) the modification to the drag force and torque on a rigid particle if it is coated by a fluid film and 2) the modification to the drag force and torque on a particle if the particle itself is confined in a drop. The former result is useful to calculate the modified Stokes drag in a variety of situations such as in describing the dynamics of ice crystals with a melted surface layer \cite{ice_drag}, particles settling through an interface \cite{coating_particle}, and core-shell microparticles in double emulsions \cite{microcapsule, control_morphology}. Similarly, the latter result is useful in understanding and describing the dynamics of confined particles, for e.g., a polymer confined in a drop \cite{polymer_conf}, encapsulated colloidal objects in drops for high through put studies of cells, bacteria, and motile organisms \cite{elegans_drop, cells_drop,microbiology_drop}, magnetic hydrogels \cite{chchen}, and encapsulated contaminants encountered in self cleaning process \cite{Wisdom}. 

Since the concentric configuration of a translating compound particle is only transient, we have also calculated the extra force required to stabilize the concentric configuration. This extra force was found to be very large for compound particles having confining drop of small thickness. 

We have analyzed the deformation dynamics of the confining drop of a compound particle when subjected to linear flows. In the limit of high surface tension (small $Ca$), we find that presence of a particle always increases the deformation of a drop. The deformation can result in breakup also, however it is possible to minimize the deformation by increasing the viscosity ratio of the outer fluid to the inner fluid and the size ratio of the drop  to the particle. A comparison between simple shear, uniaxial and biaxial flows, three typical and simple flow fields, showed that the presence of a biaxial flow is most critical for compound particles as it can rupture the confining drop most effectively. However, we have also shown that the breakup of the drop can be prevented by a pulsatile flow. In this case, the compound particle can be transported by the pulsatile flow without breaking up of the confining interface. 

Compound drops have been studied in the literature and therefore, a few of the results of compound particles are deducible from these investigations. A rotating compound particle is an exception since it has never been a subject of study previously. This is not the case for a translating compound particle. Drag force on the compound particle for a stable concentric configuration (Eq.\ref{TD-stable-dragforce}) can be deduced from the expressions obtained by \citet{Rushton} for compound drops by taking the limit of the viscosity of the fluid inclusion to infinity. Later on, \citet{Sadhal_eccentric} analyzed the compound droplet with an eccentric fluid inclusion. They used a bispherical coordinates system, from which obtaining the limit of concentric configuration is mathematically cumbersome \cite{Rushton1973}. On the other hand we have obtained closed form expressions for the drag force on a particle in both situations, namely when the compound particle attains a concentric configuration instantaneously and constantly (occurs in the absence and in the presence of an extra force on the confining drop). These calculations are useful since it reveals the role of confining interface on Stokes drag experienced by the particle as shown in Fig.~\ref{fig:TD_drag_contour_1}, the extra force required to stabilize the concentric configuration as a function of size ratio and viscosity ratio as shown in Fig.~\ref{fig:TD_stable_drag_force} and to contrast the velocity fields obtained in these two cases as shown in Fig.~\ref{fig:translation}.

Similarly, the steady state solution of a compound droplet subjected to a linear flow was known in the literature  \cite{lealstone, Brenner} but not the transient dynamics. We use the normal velocity and normal stress balance boundary conditions at the interface to calculate the unsteady evolution of the confining drop. Thus, the previous known results become deducible from our calculations. Moreover, this analysis facilitated the following: (i) A comprehensive comparison of transient dynamics of a compound particle subjected to various linear flows, namely simple shear, uniaxial and biaxial flows is discussed. (ii) In a compound particle, the surface tension of the confining interface opposes the flow driven dynamics. Conveniently, it may be thought as a competition between interface relaxation time scale and interface breakup time scale. We show that existence of these two time scales can be exploited to stably transport a compound particle using a pulsatile flow, where the flow can be periodic with a time period smaller than the breakup time to avoid the breakup of the confining drop.

Present study assumes that the Reynolds number is very small and therefore the inertia of the fluid is neglected. Unsteady inertia may become important in the calculations related to transient shape dynamics, especially in the context of pulsatile flows. In such a case, unsteady Stokes equations\cite{basset} need to be solved and this will be the subject of future investigations. Similarly a perturbation technique \cite{proudman} can be employed to quantify the effect of convective inertia.

Compound particles and their dispersions are potential soft materials since their properties can be tuned by exploiting the availability in choice of two materials, their shape and configuration. However, stability of these compound particles is a matter of concern for new experiments and investigations. In this scenario, our study will be useful in two different aspects. First, our results provide guidelines in the selection of materials to produce and manipulate the morphology of concentric compound particles. Second, the analysis of a compound particle in a linear flow helps in the design of microfluidic platforms to generate and transport the compound particles without breaking up the confining interface.

\section*{Appendix A}
In this appendix, we provide the expression for the droplet velocity and the expression for the constants used in describing the velocity and pressure fields of a translating compound particle.
\noindent 
\begin{fleqn}[0pt]
\begin{align*}
     \mathbf{V}_D = \frac{\mathbf{V}_P}{\Delta_T} \left(4+6 \alpha ^5-4 \lambda -5 \alpha ^3 \lambda +9 \alpha ^5 \lambda \right) \hspace{1.5cm}
\end{align*}
\end{fleqn}

\begin{fleqn}[0pt]
\begin{align*}
     \Delta_{T} = 4+6 \alpha ^5-8 \lambda +9 \alpha  \lambda -10 \alpha ^3 \lambda +3 \alpha ^5 \lambda +6 \alpha ^6 \lambda +4 \lambda ^2\\-9\alpha  \lambda ^2 +10 \alpha ^3 \lambda ^2 -9 \alpha ^5 \lambda ^2+4 \alpha ^6 \lambda ^2  \hspace{3.3cm} 
\end{align*}
\end{fleqn}

\begin{fleqn}[0pt]
\begin{align*}
     \mathbf{c} = \frac{\mathbf{V}_p}{\Delta_T} (4+6 \alpha ^5-8   \lambda -5 \alpha ^3 \lambda +3 \alpha ^5 \lambda+4 \lambda ^2+5 \alpha ^3 \lambda ^2\\-9 \alpha^5 \lambda ^2)
\end{align*}
\end{fleqn}

\begin{fleqn}[0pt]
\begin{align*}
     \mathbf{m} =   \frac{3 \mathbf{V}_{P}}{a^2 \Delta_T} \left(\alpha \lambda -\alpha ^3 \lambda -\alpha \lambda ^2+\alpha ^3 \lambda ^2\right)
\end{align*}
\end{fleqn}

\begin{fleqn}[0pt]
\begin{align*}
     \mathbf{g}^{(1)} =    \frac{3 \mathbf{V}_P (a \alpha)  \lambda  }{2 \Delta_T}\left(2+3 \alpha ^5-2 \lambda +2 \alpha ^5 \lambda \right)
\end{align*}
\end{fleqn}

\begin{fleqn}[0pt]
\begin{align*}
       \mathbf{d}^{(1)} =  \frac{\mathbf{V}_P (a\alpha) ^3 \lambda}{2 \Delta_T}  \left(2 \lambda -2-3 \alpha ^3-2 \alpha ^3 \lambda \right)
\end{align*}
\end{fleqn}

\begin{fleqn}[0pt]
\begin{align*}
        \mathbf{g}^{(2)} = \frac{3 \mathbf{V}_P (a  \alpha) }{2 \Delta_T} \left(2+3 \alpha ^5-2 \lambda +2 \alpha ^5 \lambda \right)
\end{align*}
\end{fleqn}

\begin{fleqn}[0pt]
\begin{align*}
       \mathbf{d}^{(2)} =  \frac{\mathbf{V}_P   (a \alpha) ^3}{2 \Delta_T} \left(-2-3 \alpha ^5+2 \lambda -5 \alpha ^3 \lambda +3 \alpha ^5 \lambda \right)
\end{align*}
\end{fleqn}

\section*{Appendix B}
In this appendix, we provide expressions for the constants used in describing the velocity and pressure fields of a compound particle in a general linear flow. 
\begin{fleqn}[0pt]
\begin{align*}
\Delta_{L} = 4+12 \alpha +24 \alpha ^2+30 \alpha ^3+30 \alpha ^4+24 \alpha ^5+12 \alpha ^6+4 \alpha ^7\\ -\lambda(4 +12 \alpha  +24 \alpha ^2 +15 \alpha ^3  -15 \alpha ^4  -24 \alpha ^5  -12 \alpha ^6 \\ -4 \alpha ^7 ) \hspace{1cm}
\end{align*}
\end{fleqn}

\begin{fleqn}[0pt]
\begin{align*}
{d}_1 = \frac{-3 \lambda \alpha ^3}{(\alpha-1) \Delta_L} \left(5+10 \alpha +8 \alpha ^2+6 \alpha ^3+4 \alpha ^4+2 \alpha ^5\right) 
\end{align*}
\end{fleqn}

\begin{fleqn}[0pt]
\begin{align*}
{d}_{2} =  \frac{-5\lambda  \alpha
^3}{6 a^2 (\alpha-1 ) \Delta_L} \left(3+6 \alpha +4 \alpha ^2+2 \alpha ^3\right) 
\end{align*}
\end{fleqn}

\begin{fleqn}[0pt]
\begin{align*}
c_1^{(1)} =  \frac{6\lambda (a\alpha) ^3}{(\alpha-1 ) \Delta_L} \left(2+4 \alpha +6 \alpha ^2+8 \alpha ^3+10 \alpha ^4+5
\alpha ^5\right) 
\end{align*}
\end{fleqn}

\begin{fleqn}[0pt]
\begin{align*}
c_3^{(1)} = \frac{-\lambda (a\alpha) ^5 }{3 (\alpha-1 )\Delta_L} \left(2+4 \alpha +6 \alpha ^2+3 \alpha ^3\right) 
\end{align*}
\end{fleqn}

\begin{fleqn}[0pt]
\begin{align*}
c_1^{(2)} = \frac{-2 a^3 \alpha ^3}{\Delta_L} \Big( 10 +30 \alpha +60  \alpha ^2 +75 \alpha ^3+75  \alpha ^4 +60  \alpha ^5 \\+30  \alpha ^6+  10  \alpha ^{7}-\lambda(4 +12 \alpha +24 \alpha ^2 +15  \alpha ^3 - 15  \alpha ^4-24  \alpha ^5 \\ -12  \alpha ^6  -4  \alpha ^{7}) \Big)  \hspace{4.5cm} 
\end{align*}
\end{fleqn}

\begin{fleqn}[0pt]
\begin{align*}
 c_3^{(2)} = \frac{a^5 \alpha ^5}{3\Delta_L}(2+6 \alpha+12 \alpha ^2+15 \alpha ^3+15 \alpha ^4+12 \alpha ^{5}+6  \alpha ^{6} \\+ 2  \alpha ^{7}) \hspace{1cm}
\end{align*}
\end{fleqn}

\begin{fleqn}[0pt]
\begin{align*}
b_1 = \frac{-1}{4 (\alpha-1) \Delta_L}\Big(32+64 \alpha+96 \alpha ^2-72 \alpha ^3-240 \alpha ^4-177 \alpha ^5\\-114\alpha ^6-76 \alpha ^7-38 \alpha ^8+\lambda(-32  -64 \alpha -96 \alpha ^2 +72 \alpha ^3 \\ +240 \alpha ^4+72 \alpha ^5 -96 \alpha ^6  -64 \alpha ^7 -32 \alpha ^8 )\Big)\hspace{3cm}
\end{align*}
\end{fleqn}

\bibliography{main} 

\end{document}